\documentclass[useAMS,usenatbib]{mn2e}

\usepackage{graphicx}
\usepackage{times}
\usepackage{natbib}

\begin{document}
 
\title[Properties of gas clumps and gas clumping factor in the ICM]{Properties of gas clumps and gas clumping factor in the intra cluster medium}
\author[F. Vazza, D. Eckert, A. Simionescu, M. Br\"{u}ggen, S. Ettori]{F. Vazza$^{1,2,3}$%
\thanks{%
 E-mail: f.vazza@jacobs-university.de}, D. Eckert$^{4}$, A. Simionescu$^{5}$, M. Br\"{u}ggen$^{1,3}$, S. Ettori$^{6,7}$\\
$^{1}$ Jacobs University Bremen, Campus Ring 1, 28759, Bremen, Germany\\
$^{2}$ INAF/Istituto di Radioastronomia, via Gobetti 101, I-40129
Bologna, Italy \\
$^{3}$ Hamburger Sternwarte, Gojenbergsweg 112, 21029 Hamburg, Germany\\
$^{4}$ ISDC Data Center for Astrophysics, Geneva, Switzerland, \\
$^{5}$ KIPAC, Stanford University, 452 Lomita Mall, Stanford, CA 94305, USA
$^{6}$ INAF -- Osservatorio Astronomico di Bologna, Via Ranzani 1, I-40127 Bologna, Italy\\
$^{7}$ INFN, Sezione di Bologna, viale Berti Pichat 6/2, I-40127 Bologna, Italy}

\date{Accepted ???. Received ???; in original form ???}
\maketitle

\begin{abstract}

The spatial distribution of gas matter inside galaxy clusters is not  
completely smooth, but
may host gas clumps associated with substructures. 
These overdense gas substructures are generally
a source of {\it unresolved} bias of X-ray observations towards high
density gas, but their bright luminosity peaks may be  {\it resolved} sources within the ICM, that deep X-ray exposures
may be (already) capable to detect.
In this paper we aim at investigating both features, using a set of high-resolution cosmological
simulations with {\small ENZO}. 
First, we monitor how the bias by unresolved gas clumping may yield incorrect estimates  
of global cluster
parameters and affects the measurements of baryon fractions by X-ray  
observations. We find that based on X-ray observations of narrow
radial strips, it is difficult  to recover the real baryon fraction to  
better than 10 - 20 percent uncertainty.
Second, we investigated the possibility of observing bright X-ray clumps in
the nearby Universe ($z \leq 0.3$). We produced simple mock X-ray  
observations for several instruments
({\it XMM}, {\it Suzaku} and {\it ROSAT}) and extracted the statistics of potentially  detectable bright clumps.
Some of the brightest clumps predicted by simulations may already have  
been already detected in X-
ray images with a large field of view. However, their small projected  
size makes it
difficult to prove their existence based on X-ray morphology only.  
Preheating, AGN feedback and cosmic
rays are found to have little impact on the statistical properties of  
gas clumps.

\end{abstract}

\label{firstpage}

\section{Introduction}
\label{sec:intro}

The properties of gas matter in more than two-thirds of the galaxy cluster volume
are still largely unknown.
In order to further improve our use of clusters as high-precision cosmological tools, it is therefore necessary to gain insight into the thermodynamics of their outer regions 
\citep[see][and references therein]{2011MSAIS..17...47E}.
The surface brightness distribution in clusters outskirts has been studied with {\it ROSAT} PSPC \citep[e.g.][and references therein]{eckert12}
and with {\it Chandra} \citep[see][and references therein]{2009A&A...496..343E}.  A step forward has been the recent observation of a handful of nearby clusters with the Japanese satellite {\it Suzaku} that, despite the relatively poor PSF and small field of view
of its X-ray imaging spectrometer (XIS), benefits from the modest background associated 
to its low-Earth orbit 
 (see results on PKS0745-191, \citealt{geo09}; A2204, \citealt{2009A&A...501..899R};
A1795, \citealt{2009PASJ...61.1117B}; A1413, \citealt{2010PASJ...62..371H}; A1689, \citealt{2010ApJ...714..423K}; Perseus, \citealt{si11}; Hydra A, \citealt{2012arXiv1203.1700S}; see also \citealt{2011arXiv1112.3030A}). 
The first results from {\it Suzaku} indicate a flattening and sometimes even an inversion of the entropy profile
moving outwards. The infall of cool gas from the large-scale structure might cause the assumption of hydrostatic equilibrium to break down in these regions, which could have important implications on cluster mass measurements \citep[][]{rasia04,lau09,bu10}. However, as a consequence of the small field of view (and of the large solid angle
covered from the bright nearby clusters) observations
along a few arbitrarily chosen directions often yield very different results.

In addition to instrumental effects \citep[e.g.][]{eck11}, or non-gravitational effects \citep[e.g.][]{ro06,lapi10,ma11,2010ApJ...725...91B,scienzo,bode12}, there are "simpler" physical
reasons to expect that properties of the ICM derived close to $\sim R_{\rm 200}$ may be more complex than what is expected from idealized cluster models.

One effect is cluster triaxiality and asymmetry, which may cause variations in
the ICM properties along 
directions, due to the presence of large-scale filaments in particular sectors of each cluster \citep[e.g.][]{va11scatter,eckert12}, or to the cluster-to-cluster variance related to the surrounding environment \citep[e.g.][]{bu10}. 

A second mechanism related to simple gravitational physics is 
gas clumping, and its possible variations across different directions from the cluster centre  \citep[][]{1999ApJ...520L..21M,nala11}. 

In general, gas clumping may constitute a source of uncertainty in the derivation of properties of galaxy clusters atmospheres since a significant part of detected photons
may come from the most clumpy structures of the ICM, which may not be fully representative
of the underlying large-scale distribution of gas \citep[e.g.][]{ro06,va11scatter}.  
Indeed, a significant fraction of the gas mass in cluster outskirts may be
in the form of dense gas clumps, as suggested by recent simulations \citep[][]{nala11}. In such clumps the emissivity of the gas is high, leading to an overestimation of the gas density if the assumption of constant density in each shell is made. The recent results
of \citet{nala11} and \citet{eckert12} show that the
treatment of gas clumping factor slightly improves the agreement
between simulations and observed X-ray profiles.

The gas clumping factor can also bias the derivation of the total hydrostatic gas mass in galaxy clusters at the $\sim 10$ percent level \citep[][]{1999ApJ...520L..21M},
and it may as well bias the projected temperature low \citep[][]{2005ApJ...618L...1R}. Theoretical models of AGN feedback also suggest that overheated clumps of gas may lead to a more efficient distribution of energy within cluster cores \citep[][]{2011MNRAS.415.2566B}.

Given that the effect of gas clumping may be {\it resolved} or {\it unresolved} by real X-ray telescopes depending on their effective resolution
and sensitivity, in this paper we aim at addressing both kind of effects, with an extensive analysis of a sample of massive ($\sim 10^{15} M_{\odot}$) galaxy clusters simulated at high spatial and mass resolution (Sec.\ref{sec:simulations}).
First, we study in Sec.\ref{subsec:clumping} the bias potentially present in cluster profiles derived without resolving the gas substructures. Second, 
in Sec.\ref{subsec:clumps} we derive the observable distribution of 
X-ray bright clumps, assuming realistic resolution and sensitivity
of several X-ray telescopes.
We test the robustness of our results with changing resolution and with additional non-gravitational processes (radiative cooling, cosmic rays, AGN feedback) in Sec.3.3.
Our discussion and conclusions are given in Sec.\ref{sec:conclusions}.

\begin{figure}
\includegraphics[width=0.45\textwidth]{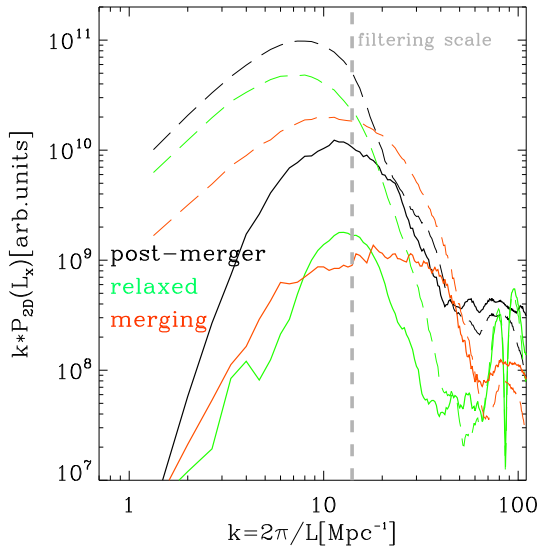}
\caption{2-dimensional power spectra of X-ray maps for three simulated clusters at $z=0$ (E1, E3B and E15B, as in Fig.\ref{fig:fig1}). Only one projection is considered. The long-dashed lines show the power spectra of the total X-ray image of each cluster, while the solid lines show the spectrum of each image after the average X-ray profile has been
removed.  Zero-padding has been considered to deal with the
non-periodic domain of each image. The spectra are given in arbitrary code units, but the relative difference in normalization of each
cluster spectrum is kept. The vertical grey line shows our filtering scale to extract X-ray clumps within each image.}
\label{fig:pks}
\end{figure}

\begin{figure*}
\includegraphics[width=0.95\textwidth]{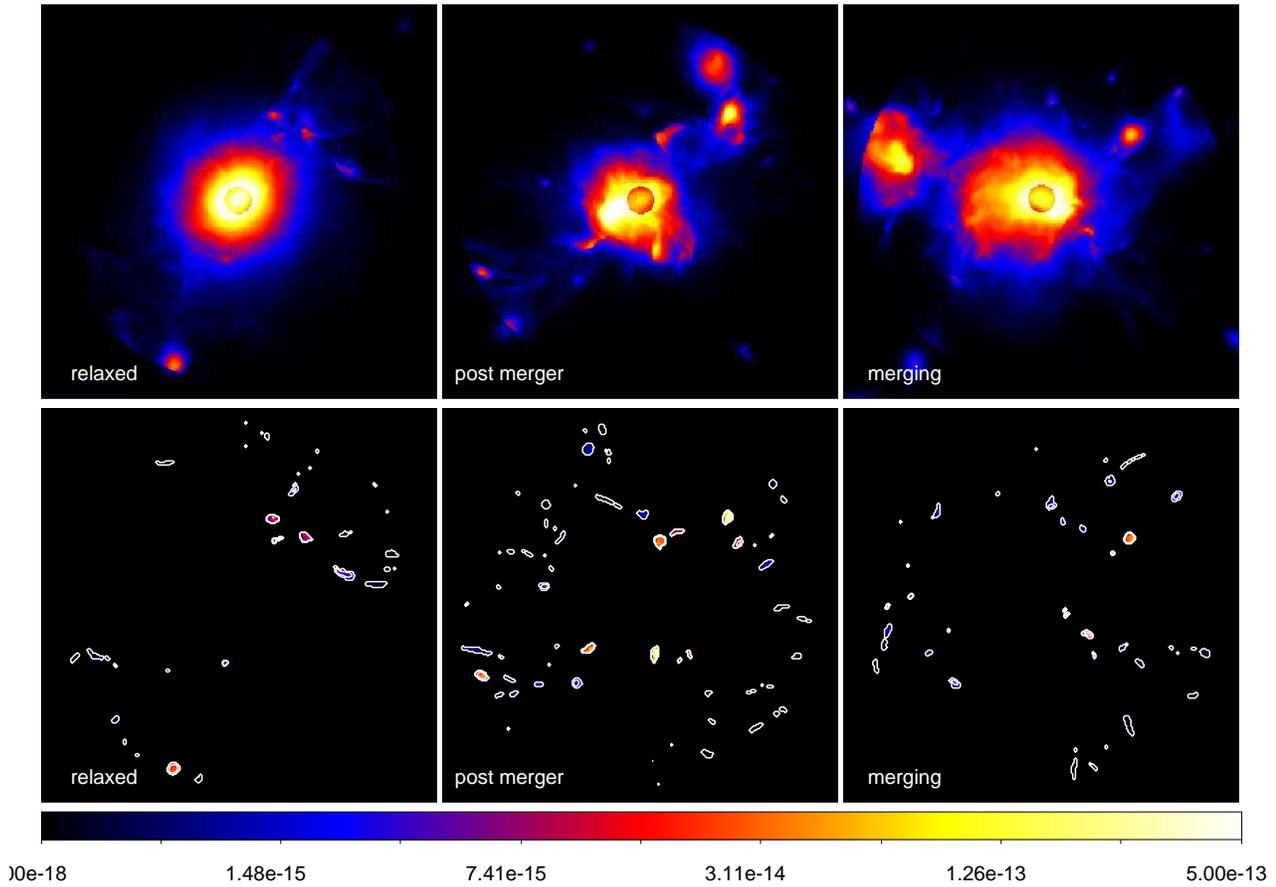}
\caption{Top panels: X-ray flux in the [0.5-2] keV (in [$\rm erg / (s \cdot cm^2)$]) of three simulated clusters of our sample at z=0 (E15B-relax, E1-post merger and E3B-merging).  Bottom panels: X-ray flux of clumps identified by our procedure (also highlighted with white contours). The inner and outer projected area excluded from our analysis have been shadowed. The area shown within each panel is $\sim 3 \times 3 ~\rm R_{200}$ for each object.} 
\label{fig:fig1}
\end{figure*}

\begin{table}
\begin{center}
\caption{Main characteristics of the simulated clusters at $z=0$. 
Column 1: identification number; 2:
total virial mass ($M_{vir}=M_{\rm DM}+M_{gas}$); 3: virial radius ($R_{v}$); 4:dynamical classification: RE=relaxing, ME=merging or  MM=major merger; 5: approximate redshift of  the last merger event.}
\begin{tabular}{c|c|c|c|c}
ID & $M_{vir}$ & $R_{v}$  & dyn.state & redshift \\
   & [$10^{15}M_{\odot}$] & [$Mpc$] & & $z_{\rm MM}$\\
\hline
E1 & 1.12  & 2.67 & MM & 0.1\\
E2 & 1.12 & 2.73 & ME  & - \\
E3A & 1.38 & 2.82 & MM & 0.2  \\
E3B & 0.76 & 2.31 & ME & - \\
E4 & 1.36 & 2.80 & MM & 0.5\\   
E5A & 0.86 & 2.39 & ME & -\\  
E5B & 0.66 & 2.18    & ME & -\\
E7 & 0.65 & 2.19    & ME & - \\ 
E11 & 1.25 & 2.72   & MM & 0.6\\ 
E14 & 1.00 & 2.60 &  RE & -\\
E15A & 1.01 & 2.63 & ME & -\\
E15B & 0.80 & 2.36 & RE & -\\
E16A & 1.92 & 3.14  & RE & -\\ 
E16B & 1.90 & 3.14  & MM & 0.2 \\ 
E18A & 1.91 & 3.14  & MM & 0.8 \\  
E18B & 1.37 & 2.80  &  MM & 0.5\\   
E18C & 0.60  & 2.08  & MM & 0.3 \\  
E21 & 0.68 & 2.18 & RE & -\\ 
E26 & 0.74 & 2.27 & MM & 0.1\\  
E62 & 1.00 & 2.50  & MM & 0.9 \\
\end{tabular}
\end{center}
\end{table}

\section{Cluster simulations}
\label{sec:simulations}

The simulations analysed in this work were produced with the  
Adaptive Mesh Refinement code {\small ENZO 1.5}, developed by the Laboratory for Computational
 Astrophysics at the University of California in San Diego 
{\footnote {http://lca.ucsd.edu}} \citep[e.g.][and references therein]{no07,co11}.
We simulated twenty galaxy clusters with masses in the range $6 \cdot 10^{14} \leq M/M_{\odot} \leq 3 \cdot 10^{15}$, extracted from a total cosmic volume of 
$L_{\rm box} \approx 480$ Mpc/h. With the use of a nested grid approach we achieved high mass and spatial resolution in the region of cluster formation: $m_{\rm dm}=6.76 \cdot 10^{8} M_{\odot}$ for the DM particles and $\sim 25 ~\rm kpc/h$ in most of the cluster volume inside the "AMR region" (i.e. $\sim 2-3 ~R_{\rm 200}$ from the cluster centre, see \citealt{va10kp,va11nice,va11turbo} for further details).

We assumed a concordance $\Lambda$CDM cosmology, with $\Omega_0 = 1.0$, $\Omega_{\rm B} = 0.0441$, $\Omega_{\rm DM} =
0.2139$, $\Omega_{\Lambda} = 0.742$, Hubble parameter $h = 0.72$ and
a normalization for the primordial density power spectrum of $\sigma_{8} = 0.8$. Most of the runs we present in this work (Sec.3.1-3.2) neglect radiative cooling, star formation and AGN feedback processes. 
In Sec.3.3, however, we discuss additional runs where the following non-gravitational processes are modelled: radiative cooling, thermal feedback from AGN, and pressure feedback from cosmic ray particles (CR) injected at cosmological shock waves.

For consistency with our previous analysis on the same sample of galaxy clusters \citep[][]{va10kp,va11turbo,va11scatter}, we divided our sample in dynamical classes based on the total matter accretion history of each halo for $z \leq 1.0$. 
First, we monitored the time evolution of the DM+gas mass for every object inside the "AMR region" in the range $0.0 \leq z \leq 1.0$. 
Considering a time lapse of $\Delta t=1~\rm Gyr$, "major merger" events are detected as 
total matter accretion episode where $M(t+\Delta t)/M(t)-1>1/3$. 
The systems with a lower accretion rate were further divided 
by measuring the ratio between the total kinetic energy of 
gas motions and the thermal energy inside the virial radius
at $z=0$, since this quantity parameter provides an indication of the dynamical 
activity of a cluster \citep[e.g.][]{TO97.2,2006MNRAS.369L..14V}. 
Using this proxy, we defined as 
"merging" systems those objects that present an energy ratio $>0.4$, but did not 
experienced a major merger in their past (e.g. they show evidence
of ongoing accretion with a companion of comparable size, but 
the cores of the two systems did not encounter yet). The remaining
systems were classified as "relaxed". 
According to the above classification scheme, our sample presents 4 relaxed
objects, 6 merging objects
and 10 post-merger objects.

Based on our further analysis of this sample, this classification 
actually mirrors a different level of dynamical activity in the 
subgroups, i.e. relaxed systems on average host weaker shocks \citep[][]{va10kp}, they are characterized by a lowest turbulent to thermal energy ratio \citep[][]{va11turbo}, and they are characterized by the smallest amount of azimuthal scatter in the
gas properties \citep[][]{va11scatter,eckert12}. 
In \citet{va11scatter} the same sample was also divided based on the
analysis of the power ratios from the multi-pole decomposition of the X-ray surface 
brightness images ($P_3/P_0$), and the centroid shift ($w$), as described by \citet{boh10}. These morphological parameters of projected X-ray emission
maps were measured inside the innermost projected $1~\rm Mpc^2$. This leads to decompose our sample into 9 "non-cool-core-like" (NCC)
systems, and 11 "cool-core-like" systems (CC), once that fiducial
thresholds for the two parameters \citep[as in][]{cassano10} are 
set. We report that (with only one exception) the NCC-like class here almost perfectly overlap with the
class of post-merger systems of \citet{va10kp}, while the CC-like class contains the relaxed and merging classes of \citet{va10kp}.

Table 1 lists of all simulated clusters, along with their main parameters measured at $z=0$.
All objects of the sample have a final total mass $> 6 \cdot 10^{14}M_{\odot}$, 12 of them having a total mass $>10^{15}M_{\odot}$. 
In the last column, we give the classification of the dynamical state of each cluster at $z=0$, and the estimated
epoch of the last major merger event (for post-merger systems).

\subsection{X-ray emission} 
\label{subsec:xray}

We simulated the X-ray flux ($S_X$) from our clusters,
assuming a single temperature plasma in ionization equilibrium within each 3D cell. We use the APEC emission model \citep[e.g.][]{2001ApJ...556L..91S} to compute the cooling function $\Lambda(T,Z)$ (where $T$ is the temperature and $Z$ the gas metallicity) inside a given energy band, including continuum and line emission.
For each cell in the simulation we assume a constant metallicity of $Z/Z_{\odot}=0.2$ (which is a good approximation of the observed metal abundance in cluster outskirts, \citealt{2008A&A...487..461L}). While line cooling may be to first approximation not very relevant for the global description of the hot ICM phase ($T \sim 10^{8} K$), it may become significant for the emission from clumps, 
because their virial temperature can be lower than that of the host cluster
by a factor $\sim 10$. Once the metallicity and
the energy band are specified, we compute for each cell the X-ray luminosity, $S_X=n_H n_e \Lambda(T,Z) dV$, where $n_H$ and $n_e$ are the number density of hydrogen and electrons, respectively, and $dV$ is the volume of the cell.

\subsection{Definition of gas clumps and gas clumping factor}
\label{subsec:definition}

Although the notion of clumps and of the gas clumping factor is often used in the
recent literature, an unambiguous definition of this is non-trivial
{\footnote{While this article was under review,  \citet{2012arXiv1210.6706Z} published a work in which they also characterize the level of inhomongeities in the simulated ICM, based on the radial median value of gas density and pressure. They study the impact of cluster triaxiality and gas clumping in the derivation of global cluster properites such like the gas mass and the gas mass fraction, reporting conclusions in qualitative agreement with our work.}}.
In this work we distinguish between {\it resolved} gas clumps (detected with a filtering of the simulated X-ray maps) and {\it unresolved}
gas clumping, which we consider as an unavoidable source of bias in the derivation of global cluster parameters from radial profiles. On the theoretical point 
of view, gas clumps represent the peaks in the distribution of the gas
clumping factor, and can be identified as single "blobs" seen in projection on the cluster atmosphere if they are bright-enough to be detected.
While resolved gas clumps are detected with a 2-dimensional filtering of the 
simulated X-ray images (based on their brightness contrast with the
smooth X-ray cluster emission), the gas clumping factor is usually
estimated in the literature within radial shells from the cluster
centre. The two approaches are not fully equivalent, and in this
paper we address both, showing that in simple non-radiative runs
they are closely related phenomena, and present similar dependence on the cluster dynamical
state.\\ 

\subsubsection{Resolved gas clumps}
\label{subsubsec:clumps_det}
We identify bright gas clumps in our cluster sample by post-processing 2-dimensional mock observations of our
clusters. We do not consider
instrumental effects of real observations (e.g. degrading
spatial resolution at the edge of the field of view of 
observation), in order to provide the estimate of the theoretical maximum 
amount of bright gas clumps in simulations. Instrumental
effects depends on the specific features of the different 
telescope, and are expected to further reduce the rate of 
detection of such X-ray clumps in real observations. 
In this section we show our
simple technique to preliminary extract gas substructures in our
projected maps, using a rather large scale ($300 ~kpc/h$). 
"X-ray bright gas clumps", in our terminology, correspond to the {\it observable} part of these
substructures, namely their small ($\leq 50 ~\rm kpc/h$) compact core, which may be detected within the host cluster atmosphere according
the effective resolution of observations (Sec.\ref{subsec:clumps}).

Based on the literature \citep[e.g.][]{do05,do09}, the most massive substructures in the ICM have a (3-dimensional) linear scale smaller than 
$<300-500 ~\rm kpc/h$. 
We investigated the typical projected size of gas substructures by computing the 2-dimensional power spectra of $S_X$ for our cluster images. In order to remove the signal from the large-scale gas atmosphere we subtracted the average 2-dimensional cluster profile from each map. We also applied a 
zero-padding to take into account
the non-periodicity of the domain (see \citealt{va11turbo} and references therein).
The results are shown in Fig.\ref{fig:pks} for three representative clusters of the sample at $z=0$: the relaxed cluster E15B, the  post-merger cluster E1 and the merging system E3B-merging (see also
Fig.\ref{fig:fig1}). 
The long-dashed profiles show the power spectra of the
X-ray image of each cluster, while the solid lines show the spectra
after the average 2-dimensional profile of each image has been
removed.  The power spectra show that most of 
the residual substructures in X-ray are characterized by
a spatial frequency of $k>10-20$, corresponding
to typical spatial scales $l_0 < 300-600 ~\rm kpc/h$, similar
to three-dimensional results \citep[][]{do09}. A dependence on the dynamical state of the host cluster is also visible
from the power spectra: the post-merger and the merging clusters have residual X-ray emission with more power 
also on smaller scales, suggesting the presence of enhanced
small-scale structures in such systems.  We will investigate this issue in more detail in the next sections.

To study the fluctuations of the X-ray flux as a result of the gas clumps we compute maps of 
residuals with respect to the X-ray emission smoothed over the scale $l_0$. The map of clumps is then computed with all pixels from the map
of the residuals, where the condition $S_X/S_{\rm X,smooth} > \alpha$ is satisfied.
By imposing $l_0=300 ~\rm kpc/h$ ($\sim 12$ cells) and $\alpha=2$,  all evident gas substructures in the projected X-ray images are captured by the algorithm. We verified that the adoption of a larger or a smaller value of $l_0$ by a factor $\sim 2$ does not affect
our final statistics (Sec.\ref{subsec:clumps}) in a significant way.
In Figure \ref{fig:fig1} we show the projected X-ray flux in the [0.5-2] keV energy range from three representative clusters of the sample (E1-post merger, E3B-merging and E15B-relaxed, in the top panels) at $z=0$, and the corresponding maps of clumps detected by our filtering procedure (lower panels). 
The visual inspection of the maps show that our filtering procedure efficiently
removes large-scale filaments around each cluster, and identifies blob-like features in the projected X-ray map. The relaxed system shows evidence
of enhanced gas clumping along its major
axis, which is aligned with the surrounding large-scale filaments. 
Indeed, although this system is a relaxed one based on its X-ray
morphology within $R_{\rm 200}/2$ (based on the measure of X-ray
morphological parameters, \citealt{cassano10}, and to its very low
value of gas turbulence, \citealt{va11turbo}), its large-scale
environment shows ongoing filamentary accretions and small satellites, which is a rather common feature even for our relaxed cluster outside $R_{\rm 500}$.
The presence of clumps in the other two systems is more evident, and they have a more symmetric distributions. We will quantify the differences of the gas clumping
factor and of the distribution of bright clumps in these systems in the
following sections.

\subsubsection{The gas clumping factor}
\label{subsubsec:clumping_factor}

The definition of the gas clumping factor, $C_{\rho}$, follows from the computation 
of the cluster density profile averaged within radial shell: 

\begin{equation}
C_{\rho}(R) \equiv \sqrt{\frac {\int_{\Omega} {\rho^{2}(R) d\Omega}} {(\int_{\Omega} {\rho(R) d\Omega})^2}},
\label{eq:clumping}
\end{equation}

\noindent where at each radial bin from the cluster centre (which we define according to the peak of the gas mass density), $R$, we compute the angular average within radial shells (with constant width of 1 cell at the highest AMR level, equivalent to $25 ~\rm kpc/h$) from the centre of clusters out to $\approx 1.5 ~\rm R_{\rm 200}$. This definition of the gas clumping factor is often used to interpret observed departures from the smooth gas density, suggested
by some observations \citep[][]{si11,nala11}. 
However, averaging within spherical shells to compute $C_{\rho}(R)$ is a procedure prone to errors in the cases of mergers or asymmetries in the cluster atmosphere, because these phenomena break the spherical symmetry in the cluster and the a spherical average might not be a good approximation. A-priori, the presence of a large gas clumping
factor (measured as in Eq.\ref{eq:clumping}) and the increased
presence of dense gas clumps may be not associated phenomena. However, as we will see in the following Sections, a larger statistics
of gas clumps and the presence of large-scale asymmetries are closely
related phenomena, which are regularly met in perturbed galaxy clusters. Indeed, the removal of the radial average of the X-ray
profile or the filtering of $\leq 300 ~\rm kpc/h$ structures 
in X-ray images in Fig.\ref{fig:fig1} visibly highlights the same
structures, because all bright gas clumps are found within
the sectors where the largest departure from the azimuthal profile
are present (i.e. due to the presence of large-scale filaments).
In the next Sections, we will explore the trend with the cluster dynamical state (and other cluster parameters) of the gas
clumping factor and of the distribution of bright clumps, and we will suggest likely observational implications of both
complementary by-products of substructures within clusters.

\begin{figure}
\includegraphics[width=0.49\textwidth]{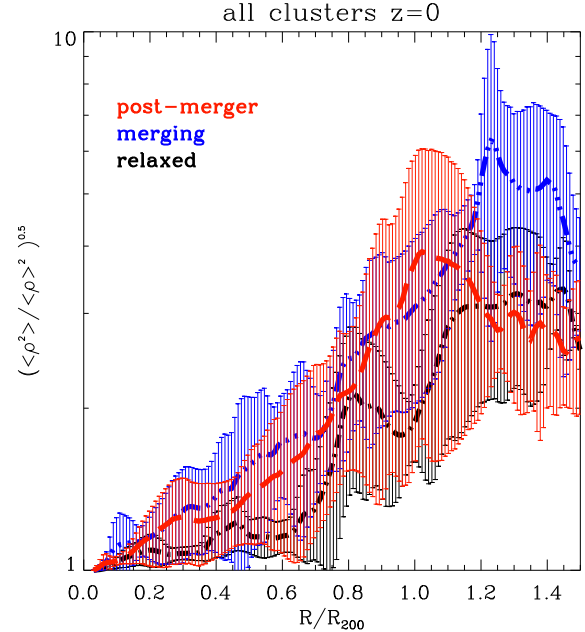}
\caption{Profile of azimuthally averaged gas clumping factors for the sample of simulated clusters at $z=0$ (the error bars show the $1\sigma$ deviation). The different colours represent the three dynamical
classes in which we divide our sample (four relaxed, six merging and ten post-merger clusters).}
\label{fig:prof_clumping1}
\end{figure}

\begin{figure*}
\includegraphics[width=0.9\textwidth,height=0.9\textheight]{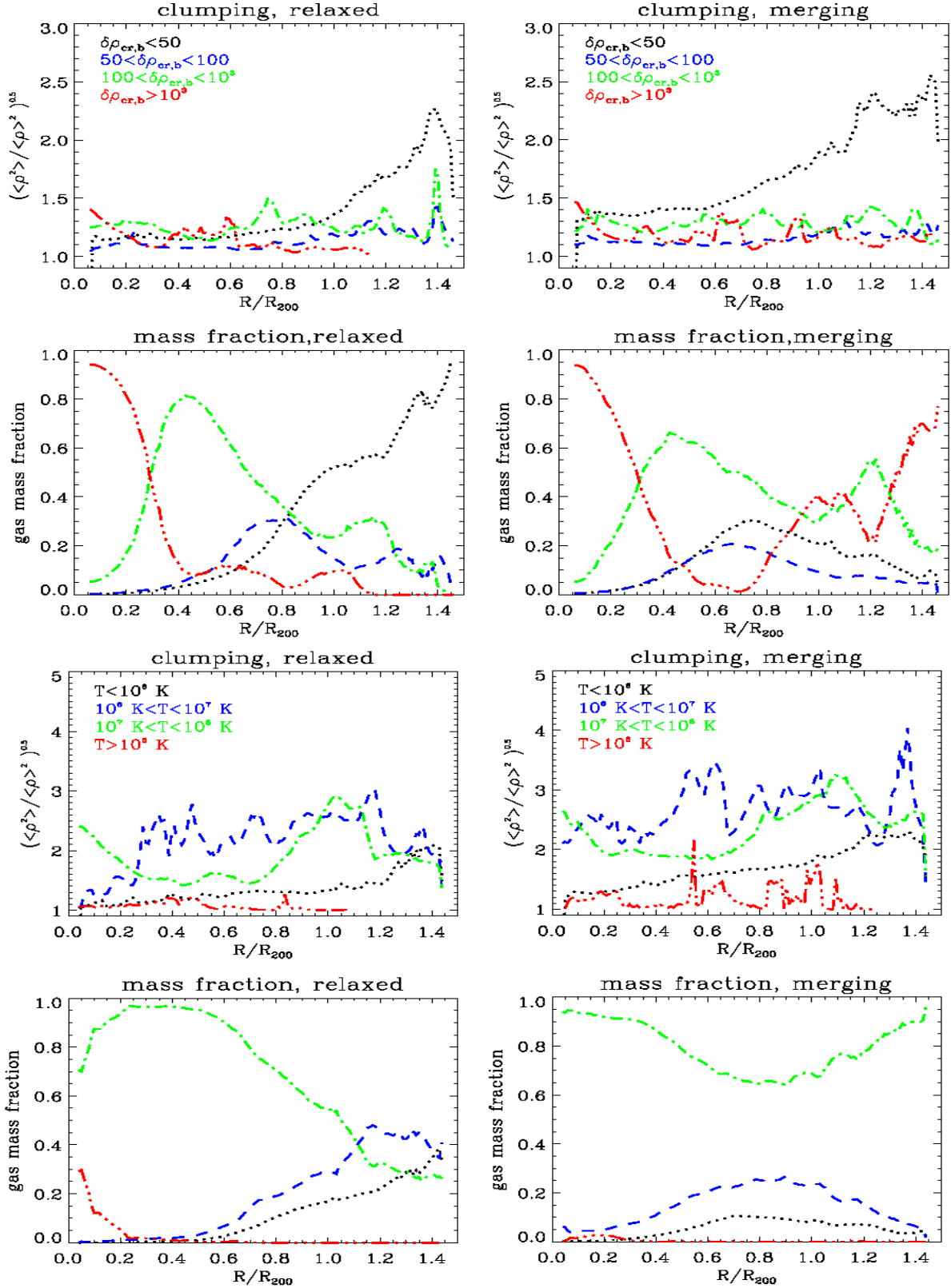}
\caption{Average  profiles  of gas clumping factor and gas mass distribution for different phases across the whole cluster sample at $z=0$, for relaxed clusters (left panels) and for merging clusters (right panels). The average gas clumping factor is computed for different decompositions of the cluster volume in gas-density bins  (top 4 panels; lines are colour-coded as described in the legend in the top 2 panels) and temperature bins (bottom 4 panels with colour-coded as detailed in paesl in the third row). The gas over-density is normalized to the cosmological critical baryon density.}
\label{fig:prof_clumping2}
\end{figure*}

\begin{figure*}
\includegraphics[width=0.33\textwidth,height=0.22\textheight]{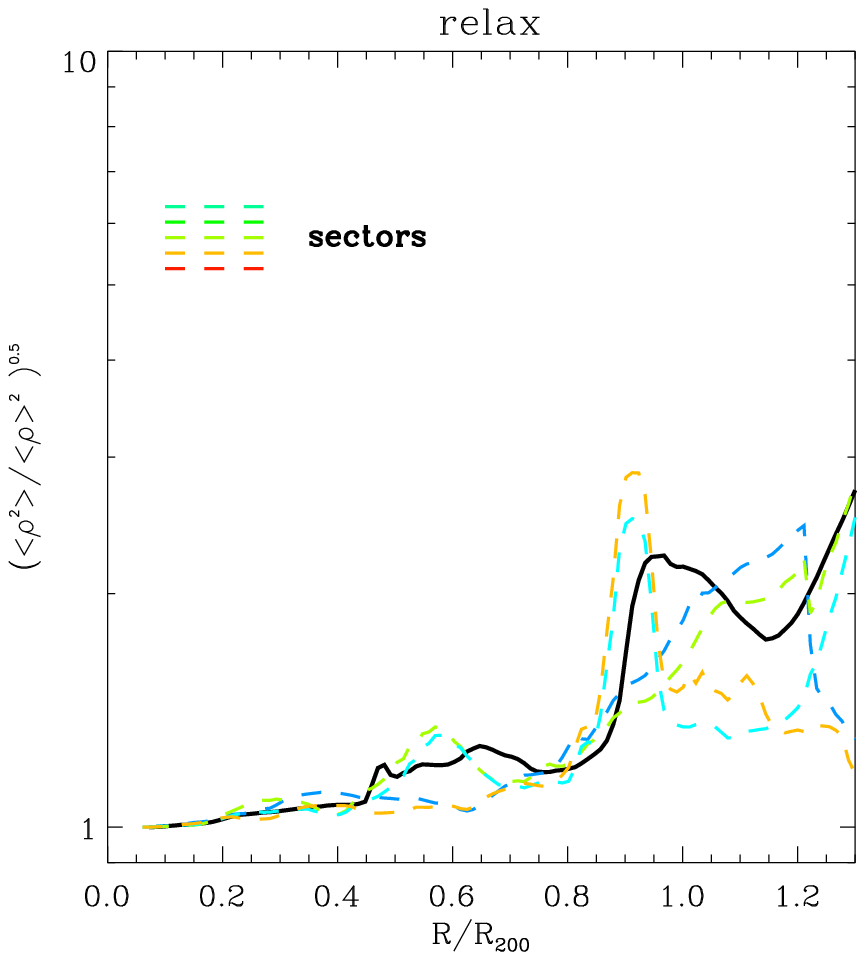}
\includegraphics[width=0.33\textwidth,height=0.22\textheight]{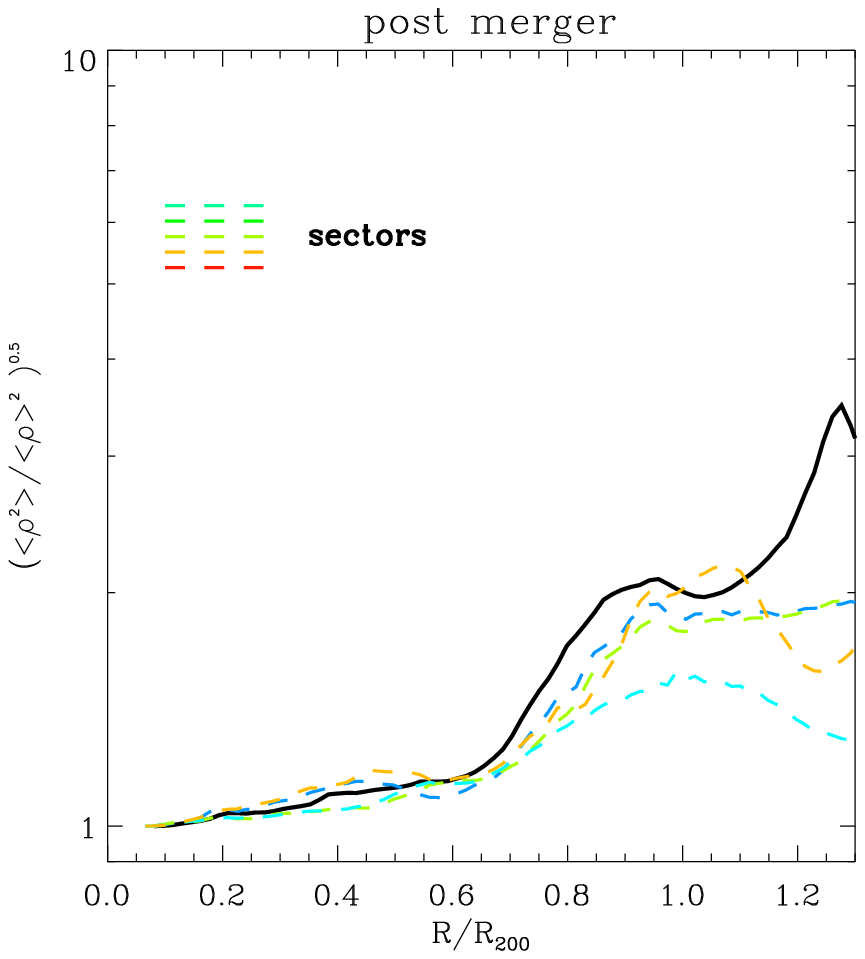}
\includegraphics[width=0.33\textwidth,height=0.22\textheight]{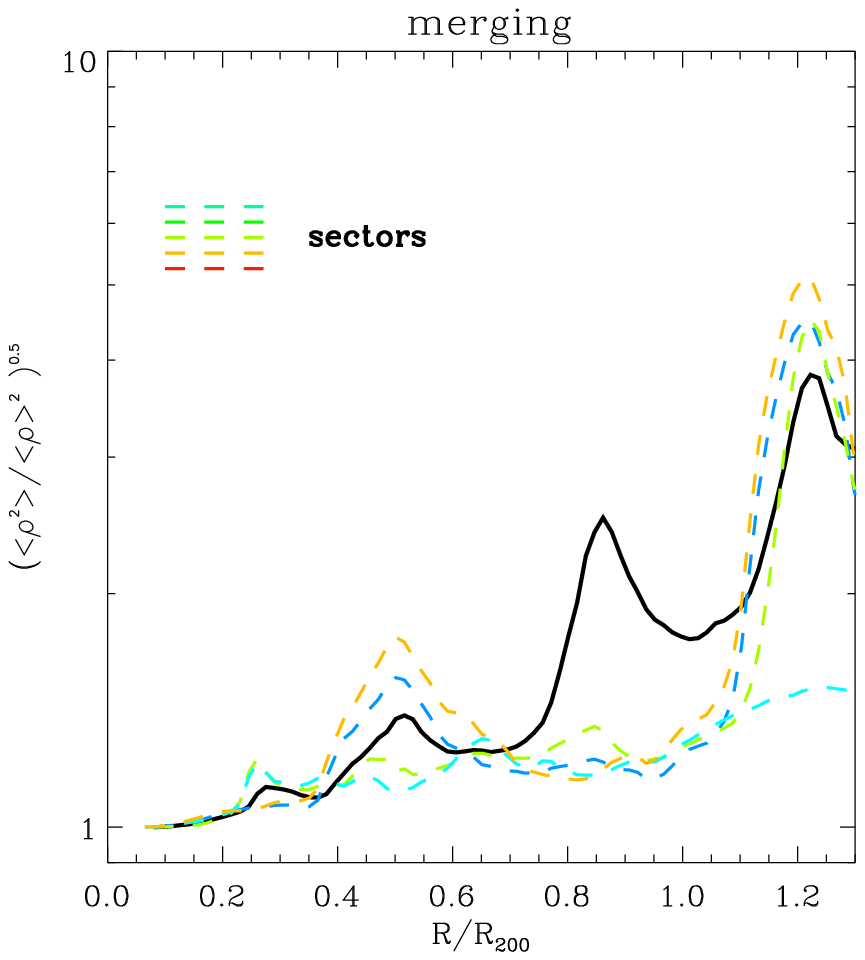}
\includegraphics[width=0.33\textwidth,height=0.22\textheight]{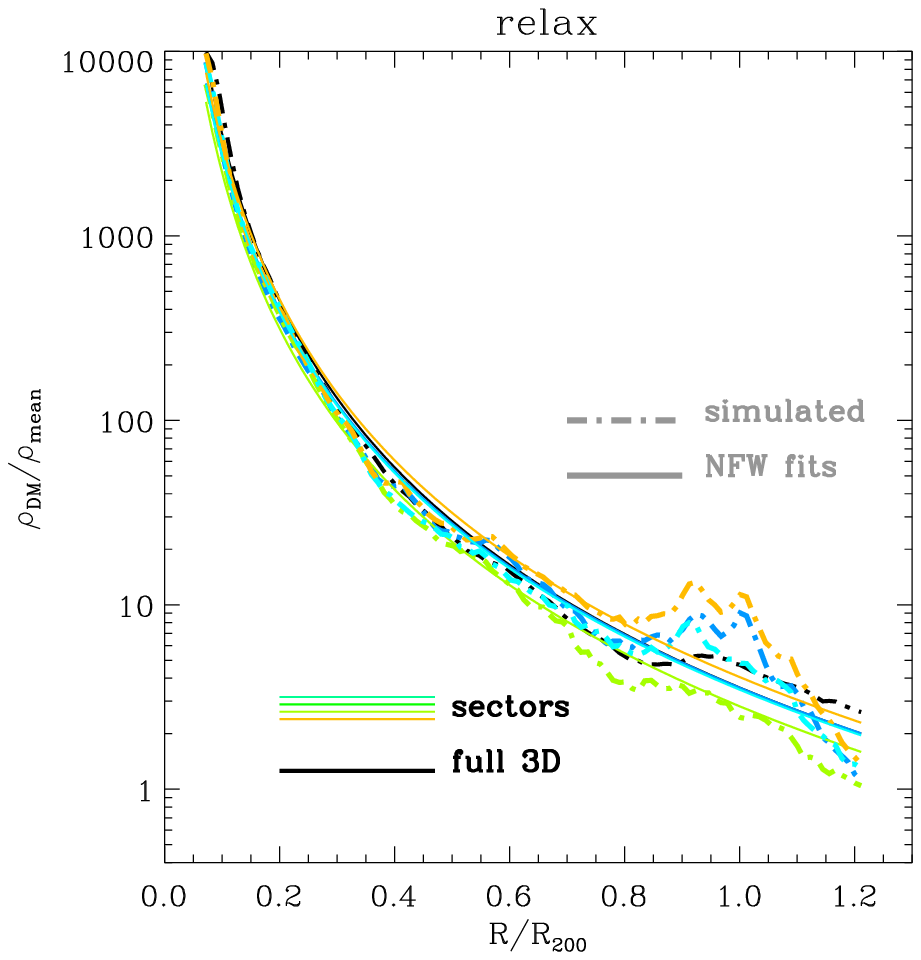}
\includegraphics[width=0.33\textwidth,height=0.22\textheight]{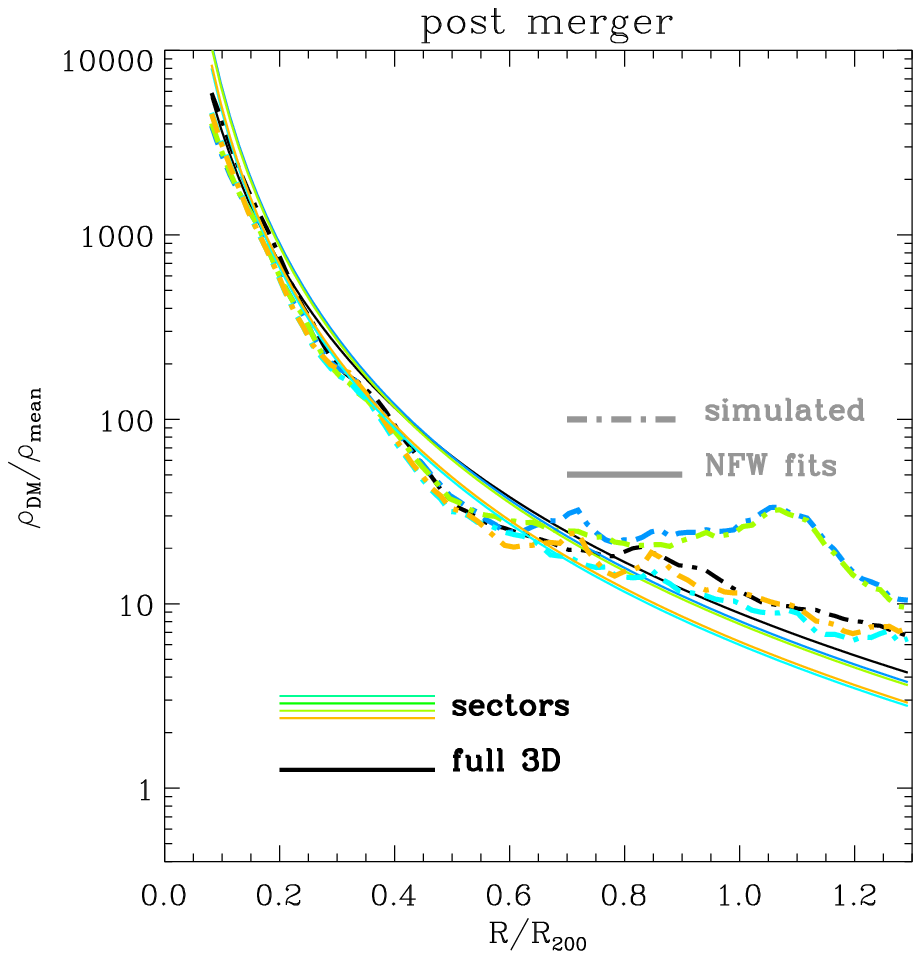}
\includegraphics[width=0.33\textwidth,height=0.22\textheight]{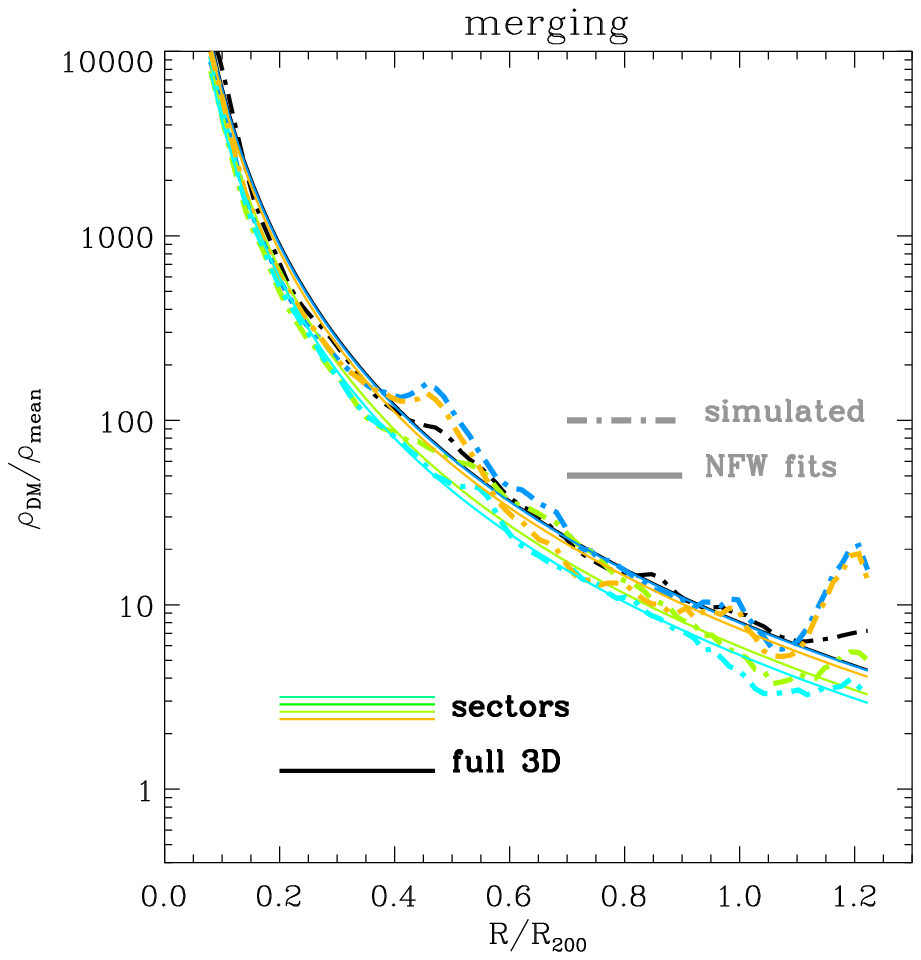}
\includegraphics[width=0.33\textwidth,height=0.22\textheight]{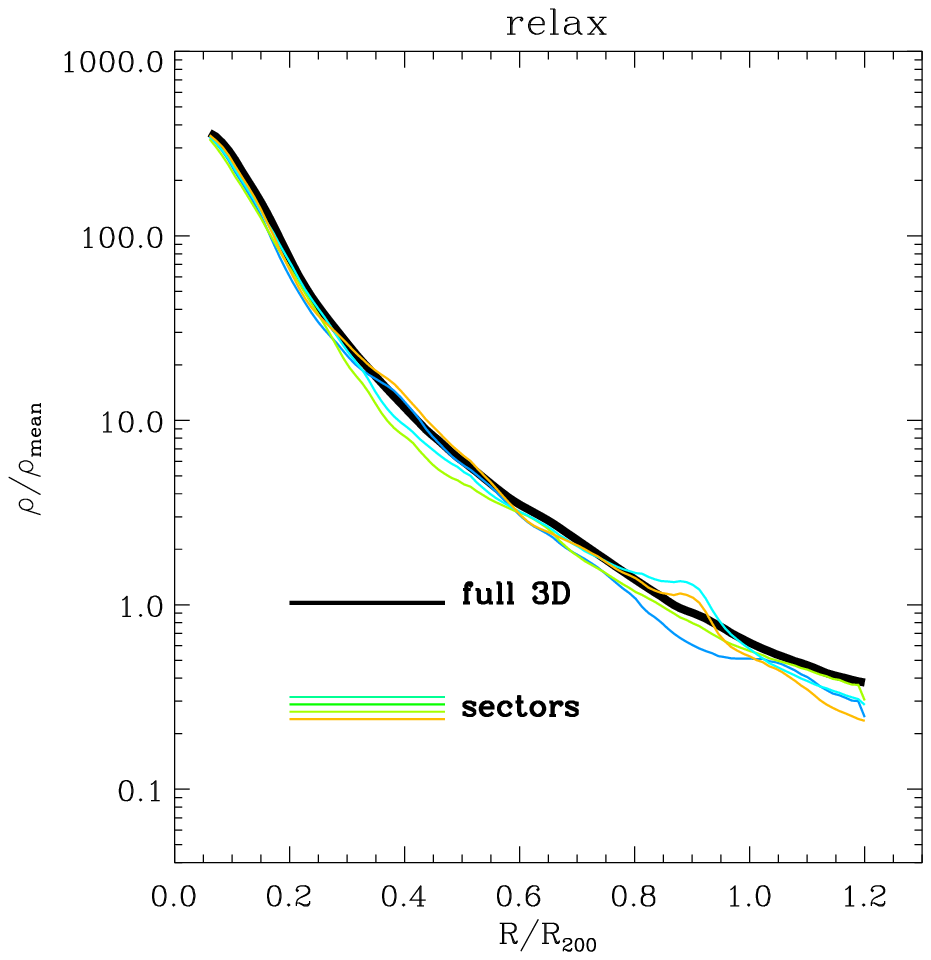}
\includegraphics[width=0.33\textwidth,height=0.22\textheight]{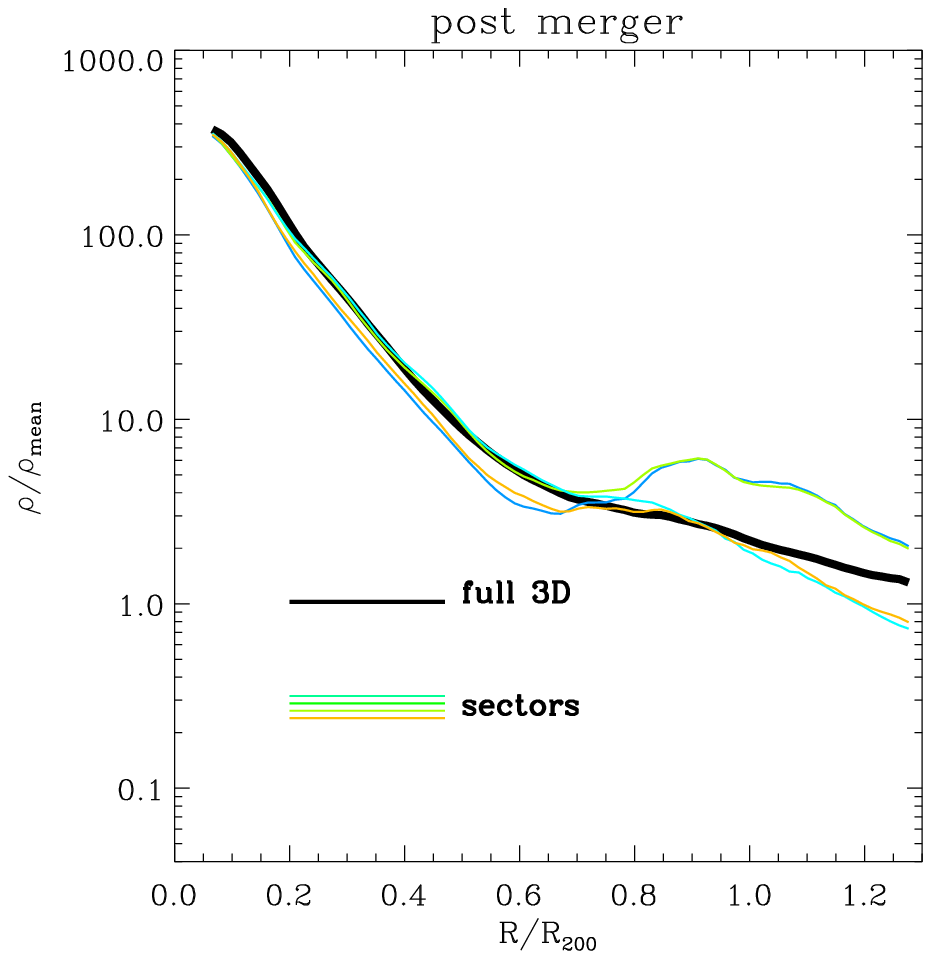}
\includegraphics[width=0.33\textwidth,height=0.22\textheight]{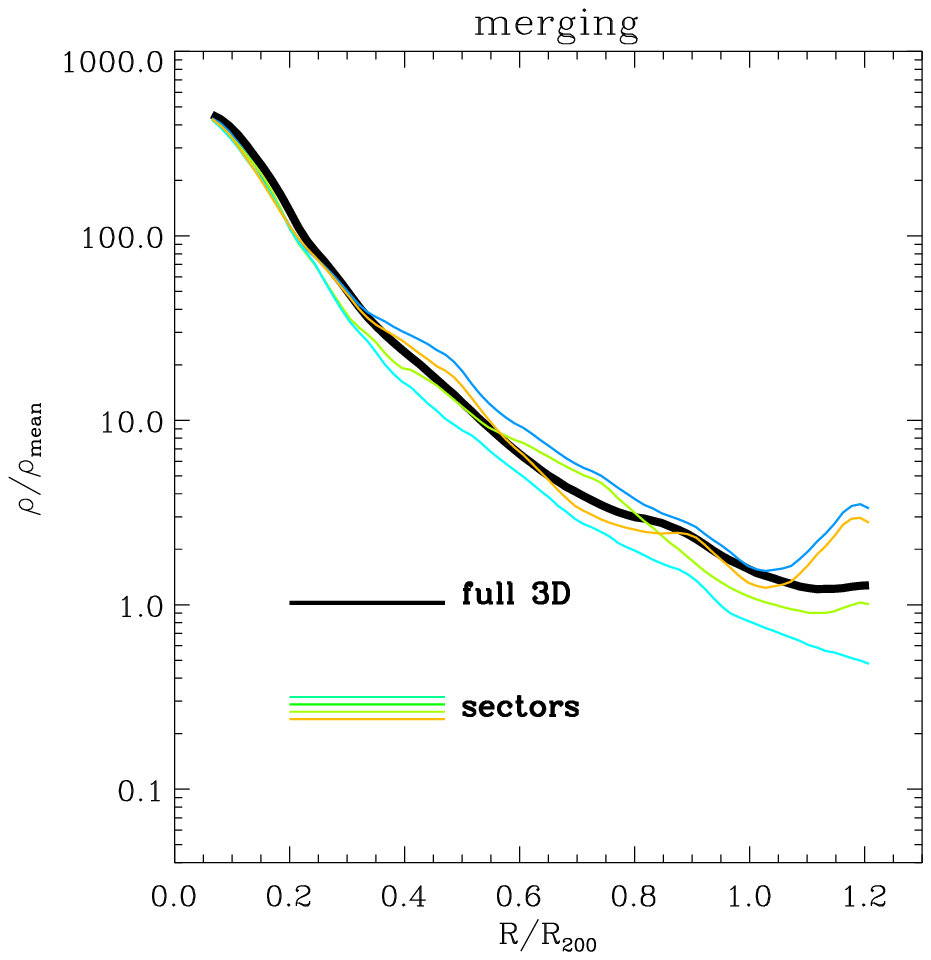}
\includegraphics[width=0.33\textwidth,height=0.22\textheight]{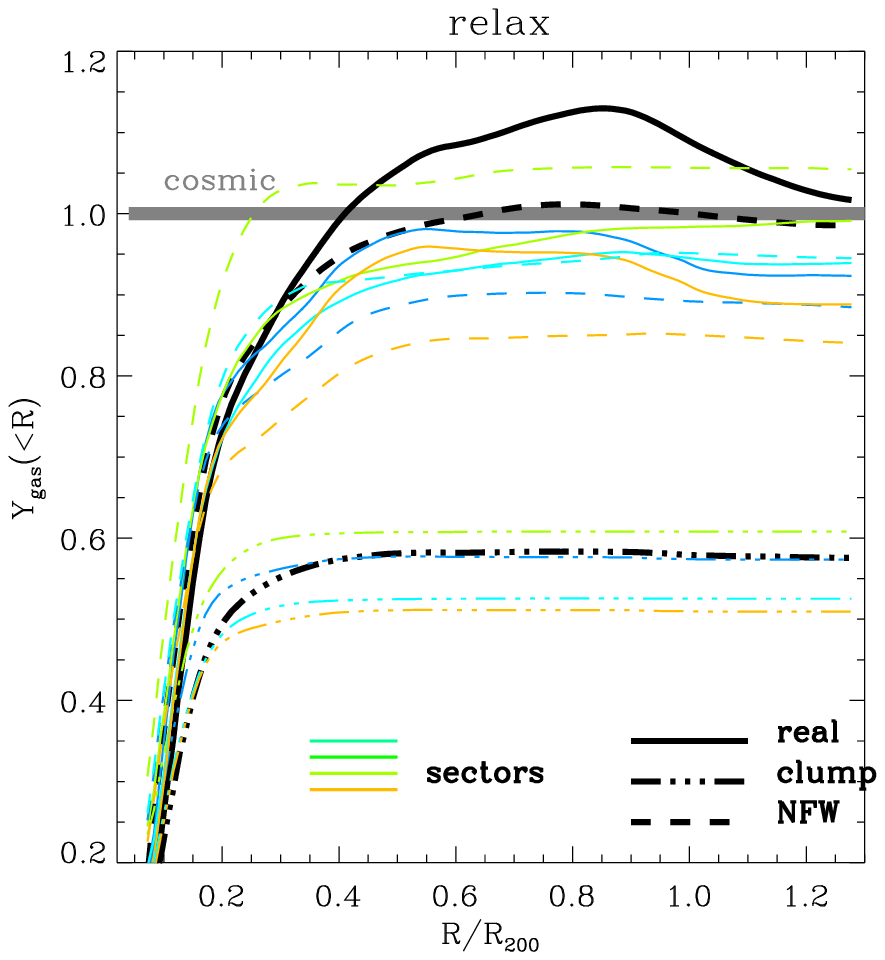}
\includegraphics[width=0.33\textwidth,height=0.22\textheight]{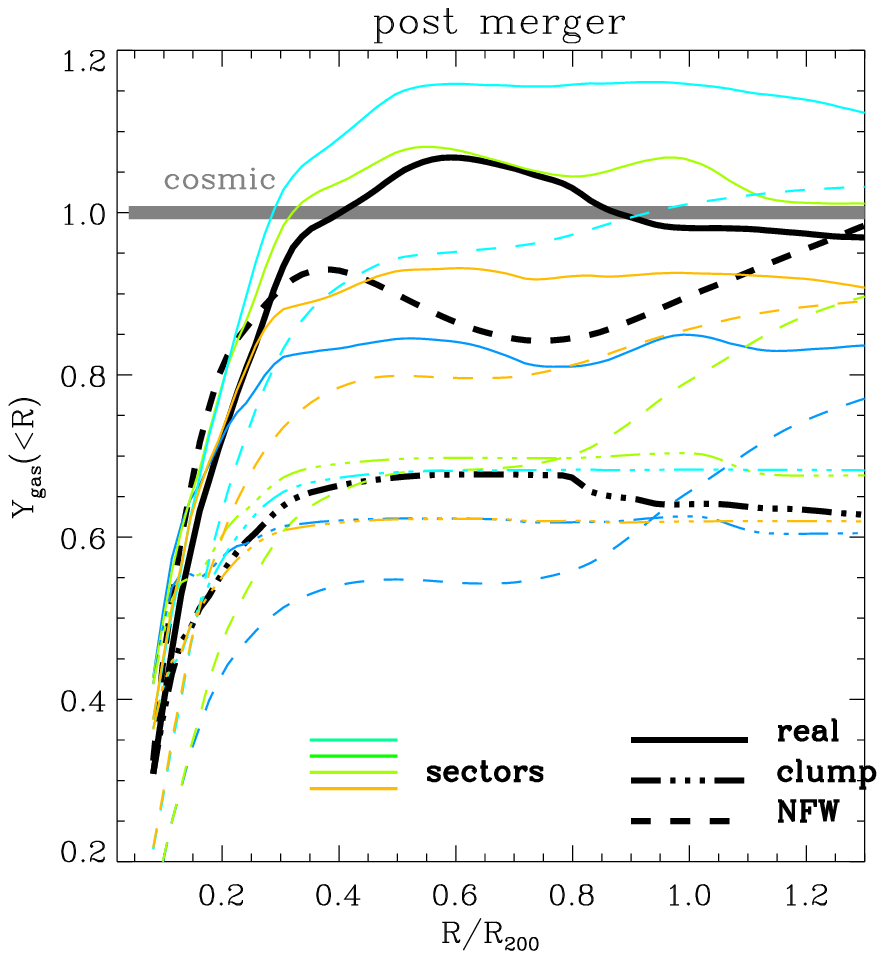}
\includegraphics[width=0.33\textwidth,height=0.22\textheight]{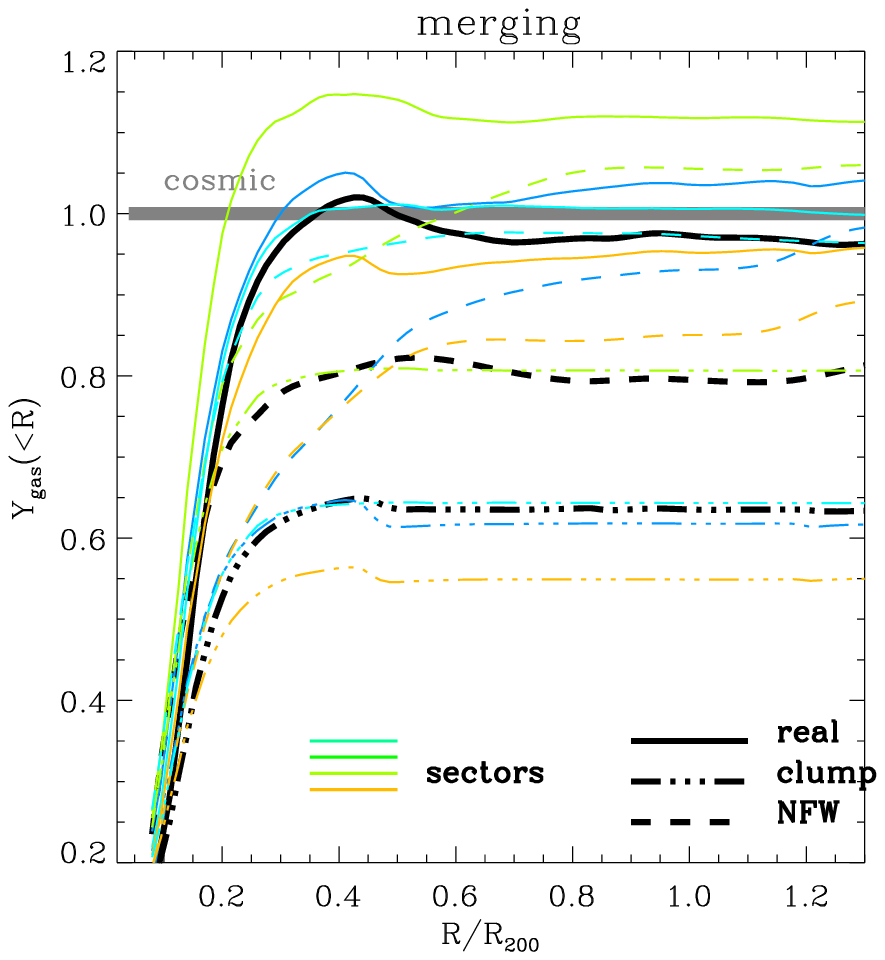}
\caption{{\it Top panels}: radial profiles of clumping factor for three clusters of the sample at $z=0$ (from left to right: E15B-relax, E1-post merger and E3B-merging). The solid black line show the average gas clumping factor for the whole volume, the additional coloured lines show the gas clumping factor inside 4 smaller sectors from the cluster centre. {\it Second row}: radial profiles of DM density for the total volume (black) and for the sectors (colors). The dot-dashed lines show the simulated data, the solid ones the best-fit of the NFW profile for the corresponding volumes. 
{\it Third row}: average gas density profiles for the same volumes (same meaning
of colours). {\it Last row}: profiles of enclosed gas mass fraction for the same volumes. We show the real profile (solid lines), the "clumpy" baryon fraction (dot-dash) and the "X-ray" baryon fraction (dashed) for each corresponding volume.}
\label{fig:prof_clumping3}
\end{figure*}

\begin{figure}
\includegraphics[width=0.4\textwidth]{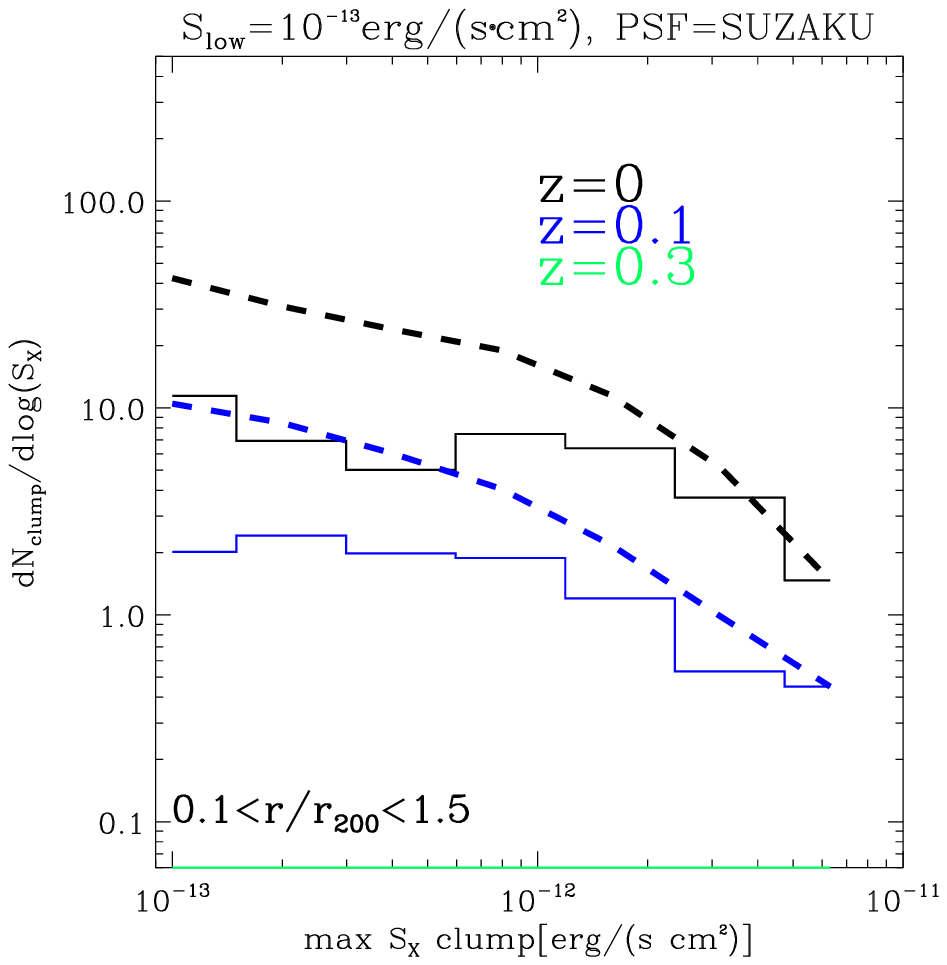}
\includegraphics[width=0.4\textwidth]{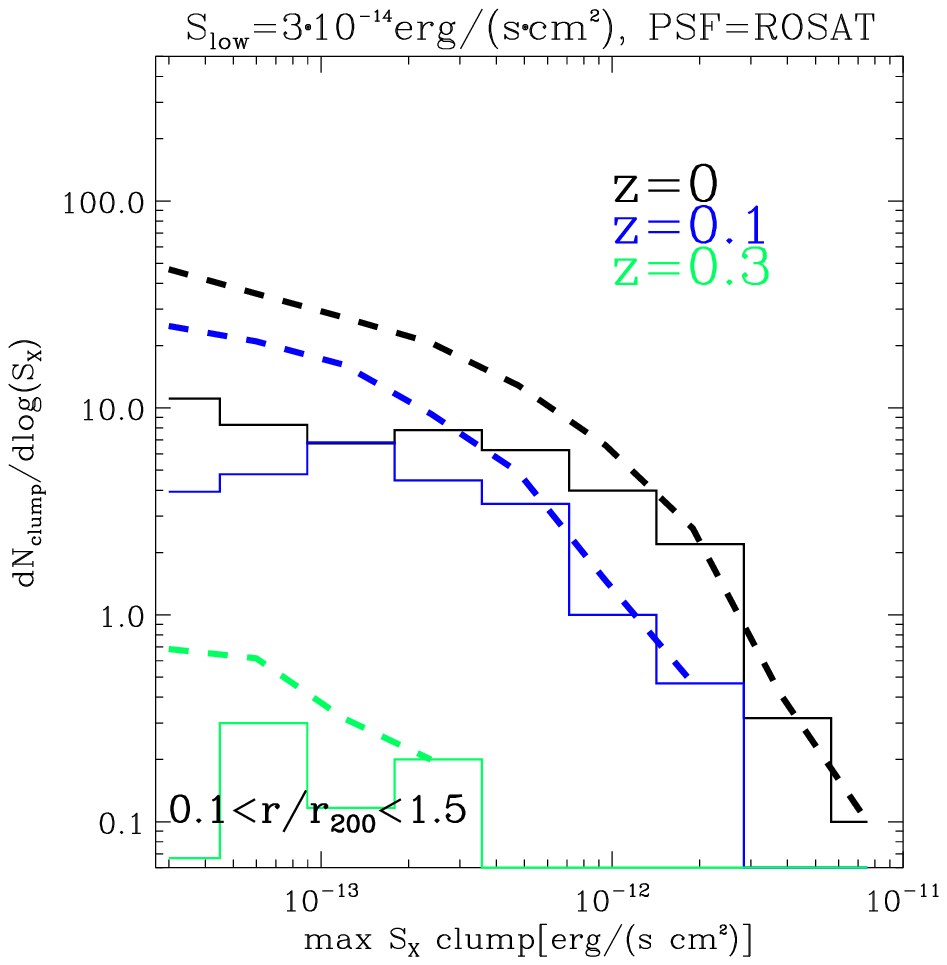}
\includegraphics[width=0.4\textwidth]{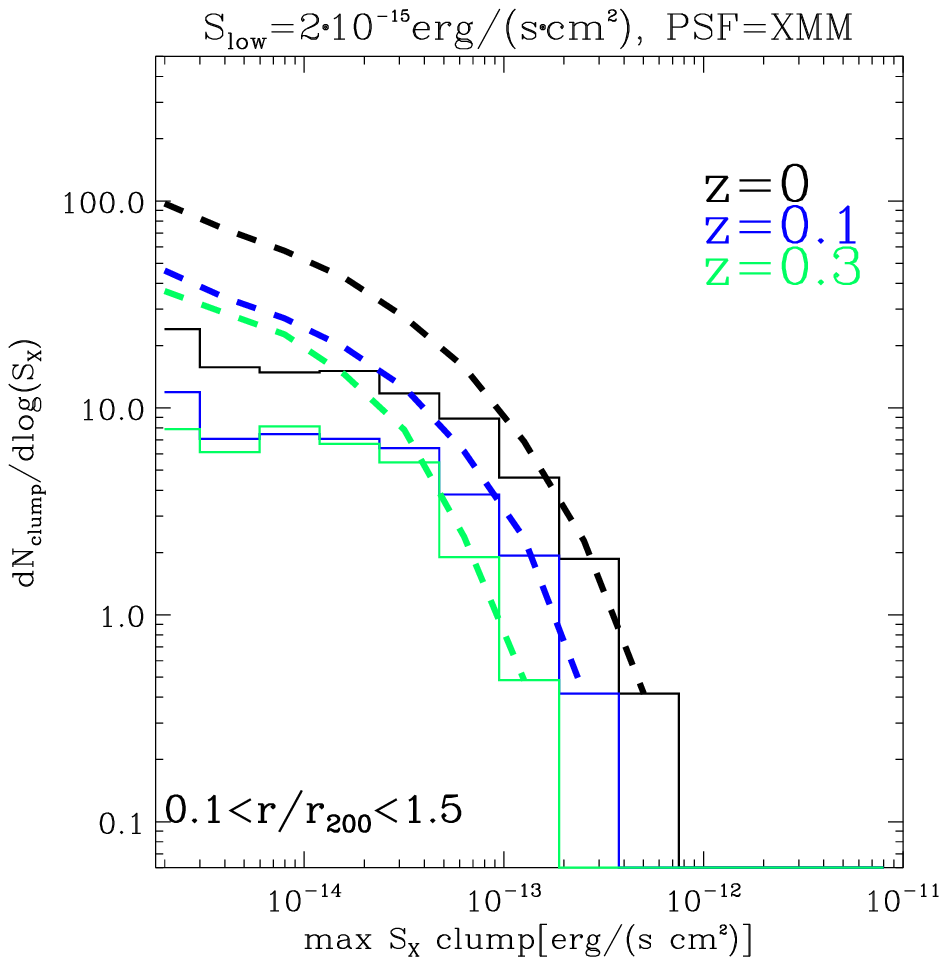}
\caption{Average luminosity function of clumps detected in our sample of simulated X-ray maps at z=0, z=0.1 and z=0.3. Each panel shows the estimate
for a different assumed X-ray telescope (from top to bottom: {\it Suzaku}, {\it ROSAT} and {\it XMM}). The continuous lines are for the differential distribution, the dashed lines are for the cumulative distributions.}
\label{fig:distrib_clumps1}
\end{figure}

\section{Results}
\label{sec:results}

\subsection{Gas clumping factor from large-scale structures}
\label{subsec:clumping}

The average clumping factor of gas in clusters is
a parameter adopted in the
theoretical interpretation of de-projected
profiles of gas mass, temperature and
entropy \citep[e.g.][]{ur11,si11,eckert12,2012MNRAS.tmp.2785W}.

For each simulated cluster, we compute the profile of the gas
clumping factor following Sec.\ref{subsubsec:clumping_factor}.

Figure \ref{fig:prof_clumping1} shows the average profile of $C_{\rho}(R)$ for the three dynamical classes into which our sample can be divided.
The trends for all classes are qualitatively similar, with the average
clumping factor $C_{\rho} < 2$ across most of the cluster volume,
and increasing to larger values ($C_{\rho} \sim 3-5$) outside $R_{\rm 200}$. Post-merger and merging systems present a systematically larger average value of gas clumping factor at all radii, and also a larger variance within the subset. These findings are consistent with the recent results of \citet{nala11}. By extracting the distribution of bright gas clumps in projected cluster images, 
we will see in Sec.\ref{subsec:clumps} that both the radial distribution of clumps, and the trend of their number with 
the cluster dynamical state present identical behaviour of the gas clumping factor measured here. This again support the idea
that the two phenomena are closely associated mechanism, following the injection of large-scale substructures in the ICM.

Next, we calculate the average profiles of gas clumping factor for different
bins of gas over-density and temperature (limiting to the relaxed and the
merging subset of clusters).  This way we can determine if the gas clumping factor affects all phases of the ICM equally. 

In Fig.\ref{fig:prof_clumping2} we show the average profiles of gas clumping factor calculated within different bins of 
gas over-density ($\delta \rho_{\rm cr,b}<50$, $50 \leq \delta \rho_{\rm cr,b} < 10^{2}$, $10^2 \leq \delta \rho_{\rm cr,b} < 10^3$ and $\delta \rho_{\rm cr,b}\geq 10^3$, where $\delta \rho_{\rm cr,b}$ is $\rho/(f_{\rm b}\rho_{\rm cr}$), with $f_{\rm b}$ cosmic baryon fraction and $\rho_{\rm cr}$ the cosmological critical density) and gas temperature ($T<10^6$ K, $10^6 \leq T < 10^7$ K, $10^7 \leq T < 10^8$ K and $T \geq 10^8$ K) at $z=0$.

We want to have a characterization of the environment associated
with the most significant source of gas clumping (e.g. gas substructures) in a way which is unbiased by their presence. This is
non-trivial, because if the local overdensity is computed within
a scale smaller than the typical size of substructures ($\leq 300 ~\rm kpc/h$), the gas density of a substructure will bias the estimate
of local overdensity high. For this reason, the local gas
overdensity and gas temperature in Fig.\ref{fig:prof_clumping2} have
bee computed for a much larger scale, $1 ~\rm Mpc/h$.{\footnote {We checked that the use of the DM over density statistics, in this case, yield very similar results to those obtained using the gas over density. However, we focus on the latter in this work, since this quantity can be more easily related to observations}}. In Fig.\ref{fig:prof_clumping2} we also show the gas mass fraction of each phase inside the  radial shells.
The gas phases show evidence that the increased clumping factor of merging clusters is related to the increased clumping factor of the low-density phase, at all radii, even if the mass of this component is not dominant in merging systems. An increased clumping factor is also statistically found in the cold ($<10^6$ K) and intermediate $10^6 \leq T < 10^7$ phase of gas in merging systems.
We remark that although the increase in this phase is associated to an enhanced presence of X-ray detectable clumps (Sec.\ref{subsec:clumps}), it cannot be directly detected in X-ray, due to 
its low emission temperature.

The gas phase characterized by $\delta \rho_{\rm cr,b}<50$ and $T \leq 10^6-10^7$ is typical of large-scale filaments
\citep[][]{do08,va09shocks,iapichino11}.  
While filaments only host a small share of the gas mass inside the virial radius, they can dominate the clumping factor within the entire volume of the merging systems.
This is because their gas matter content contains significantly more clumpy matter compared to the ICM onto which they accrete. 
This suggests that a significant
amount of gas substructure (e.g. including massive and bright X-ray clumps, as in Sec.\ref{subsec:clumps}) can originate from low-density regions, and not necessarily from the hot and dense phase of the ICM.
This is reasonable in the framework of hierarchical structure formation, since matter inside filaments had almost the entire cosmic evolution
to develop substructures \citep[][]{2011A&A...531A..75E}. Contrary to most of the matter inside galaxy clusters, large-scale filaments have never undergone
a phase of violent relaxation and efficient mixing. Therefore, even at late cosmic epochs they can supply very 
clumpy material.   
This suggests that clusters in an early merging phase are characterized by the largest amount of gas clumping factor at all radii, and that this clumpy materials is generally associated with substructures coming from filaments and from the cluster outskirts. Filaments between
massive galaxy clusters are indeed found to anticipate major mergers in
simulations \citep[][]{va11turbo}, starting the injection of turbulence
in the ICM already some $\sim \rm Gyr$ before the cores collision.
\bigskip

We now investigate how gas clumping can affect X-ray observations. 
For this purpose, we compare in Fig.\ref{fig:prof_clumping3}  the real profiles of clumping factor, DM, gas density and baryon fraction for the same clusters as in Fig.\ref{fig:fig1}, with profiles derived from 
the entire cluster volume or within smaller volumes (i.e. thin slices along the line of sight) along perpendicular sectors of the same systems.  All profiles are drawn from the centre of total (gas+DM) mass of each system.
These selected volumes are meant to broadly compare with the selection
of sectors chosen in recent deep X-ray observations of clusters 
\citep{si11,ur11,2012ApJ...748...11H}. Therefore, in this case we studied the radial profiles inside rectangular sectors of width $\sim 500 ~\rm kpc$, similar to
observations \citep[][]{si11,ur11}, and the largest available line of sight ($\sim 8 ~\rm  Mpc$). The projected total volume sampled by these sectors is $\sim 20$ percent of the total cluster volume, and it is interesting to study how representative the information is that can be derived from them.

For comparison with the observationally derived baryon fraction, we compute the best-fit model for NFW profiles \citep[][]{NA96.1} using the {\small CURVEFIT} task in {\small IDL} for the total
volume, and for each sector independently. The radial range used to compute the best-fit to the NFW profile is $0.02 ~R_{\rm 200} \leq R \leq R_{\rm 200}$.

In the relaxed system (first panels) there is little gas clumping out to a radius $0.8 R_{\rm 200}$. The average profile of DM is overall well modelled
by the NFW profile everywhere up to 
$\sim R_{\rm 200}$. The profiles inside smaller sectors are also well
described by a NFW profile, yet for $>0.6 R_{\rm 200}$ the NFW profiles
within each sector start to deviate from one another. At large radii the best-fit
profiles are found to be both under- or over-estimating the real DM profile,
due to large-scale asymmetries in the cluster atmosphere (e.g. left panel of Fig.\ref{fig:fig1}). Asymmetries in this relaxed cluster also cause a similar
azimuthal scatter of gas profiles, similar to what was found in \citet[][]{va11scatter}.

The departures from the NFW profile and the differences between sectors
are significantly  larger in the post-merger and the merging systems (second and third columns of Fig.\ref{fig:prof_clumping3}). 
We find that the best-fit NFW profiles can differ significantly between sectors, with a relative difference of up to $\sim 30-40$ percent at $R_{\rm 200}$ in the merging system caused by the infalling second cluster (see right panel of Fig.\ref{fig:fig1}).
In addition to azimuthal variations, also large 
differences ($>40$ percent) between the true DM profile and the best-fit NFW profile are found along some sectors.

These uncertainties in the "true" gas/DM mass within selected sectors
bias the estimated enclosed baryon fraction. This is an 
important effect which may provide the solution to understand the variety of recent
results provided by observations \citep[][]{si11,ur11,2012ApJ...748...11H,2012MNRAS.tmp.2785W}.

In Fig.\ref{fig:prof_clumping3}, we present the radial trend of the
enclosed baryon fraction, $Y_{\rm gas}(<R)\equiv f_{\rm bar}(<R)/f_{\rm b}$ for each
cluster and sector (where $f_{\rm b}$ is the cosmic baryon fraction), following
 three different approaches:  

\begin{itemize}
\item we measure the "true" baryon fraction inside spherical shells (or portion of spherical shells, for the narrow sectors);
\item we measure the "X-ray-like"
baryon fraction given by $(\int 4 \pi R^2 \rho(R)^2 dR)^{0.5} / \int 4 \pi R^2 \rho(R)_{\rm NFW} dR$, where $\rho_{\rm NFW}$ is the profile
given by the best-fit NFW profile for DM (for the total volume or for each sector separately); 
\item we measure the "clumpy" baryon fraction, $(\int 4 \pi R^2 \rho(R)^2 dR / \int 4 \pi R^2 \rho(R)_{\rm DM}^2 dR)^{0.5}$, where $\rho(R)_{\rm DM}$ is the true DM density). 
\end{itemize}

This "clumpy" estimate of $Y_{\rm gas}$
obviously has no corresponding observable, since it involves the clumping factor of the DM distribution. However, this measure is useful because 
it is more connected to the intrinsic baryon fraction of the massive substructures within galaxy clusters.

\bigskip

The radial trend of the true baryon fraction for the total volume is in line with the result of observations \citep[][]{2002MNRAS.334L..11A,2006ApJ...640..691V,2009A&A...501...61E}, with $Y_{\rm g}$ increasing from very low central values up to the cosmic value close to $R_{\rm 200}$. The total enclosed baryon fraction inside $<1.2 ~R_{\rm 200}$ is $\pm 5$ percent of the cosmic value, yet larger departures ($\pm 10-20$ percent) can be found inside the narrow sectors. 

The "X-ray-like" measurement of the enclosed baryon fraction is subject to a larger azimuthal scatter, and shows larger differences also with respect to the true enclosed baryon fraction. In our data-sample, it
rarely provides an agreement better than $\pm 10$ percent with the true enclosed baryon fraction (considering the whole cluster or the sectors).
Based on our sample, we estimate that on average in $\sim 25$ percent of cases the baryon fraction measured within sectors by assuming an underlying NFW profile overestimates the true baryon fraction by a $\sim 10$ percent, while in $\sim 50$ percent of cases it underestimates the cosmic baryon fraction by a slightly larger amount ($\sim 10-20$ percent). In the
remaining cases the agreement at $R_{\rm 200}$ is of the order of $\sim 5$ percent.

In all cases the main factor for significant differences between
the X-ray-derived baryon fraction and the true one is the triaxiality of
the DM matter distribution within the cluster 
rather than enhanced gas clumping. As an example,  
in the relaxed cluster of Fig.\ref{fig:prof_clumping3}
the "X-ray" baryon fraction is underestimated
in 2 sectors, it is overestimated in 1 sector and it almost follows the true one in the last sector. This trend is mainly driven
by the difference between the true DM profiles and the best-fit NFW profiles within sectors for $>0.4 ~R_{\rm 200}$, rather than dramatic differences in the gas clumping factor. 

In general, a significant gas clumping factor ($C_{\rho} \sim 2-3$) in the cluster outskirts can still lead to fairly accurate baryon fractions provided that this clumping is associated with DM substructures or DM filaments. In a realistic observation, the presence of a large-scale filament towards the cluster centre would still yield a 
reasonable best-fit NFW profile, even if not necessarily in agreement with the global NFW profile
of the cluster.  As a result, the measurement of a baryon fraction close to the cosmic value along one sector does not imply
the absence of gas clumping.
On the other hand, the presence of significant large-scale asymmetries can be responsible for strong differences between the baryon fraction measured in X-rays
and the true baryon fraction, even without substantial gas clumping factor.
To summarize, the result of our simulations is that in many realistic configurations obtained in large-scale structures, the gas clumping factor inside clusters and the enclosed baryon fraction may provide even unrelated information.
   
Also the radial range selected to fit the 
NFW profile to the observed data can significantly affect the estimate
of the enclosed baryon fraction. Given the radial gas and DM patterns
found in our simulated clusters, all radii outside $>0.4 R_{\rm 200}$ may
be characterized by equally positive or negative fluctuations compared 
to the NFW profile. Adopting a different outer radius to fit a NFW
profile to the available observations may therefore cause 
a small but not negligible ($\sim 5-10$ percent in the normalization
at the outer radii) source of uncertainty in the final estimate of the
enclosed baryon fraction.

These findings might explain why the enclosed baryon fraction is estimated
to be {\it larger} than the cosmic baryon fraction at $\sim R_{\rm 200}$, for some nearby clusters \citep[][]{si11}, {\it smaller} than the cosmic
value in some other cases \citep[][]{2010ApJ...714..423K}, and compatible with the cosmic value in the remaining cases \citep[e.g.][]{2012ApJ...748...11H,2012MNRAS.tmp.2785W}. These observations could generally only use a few selected narrow sectors, similar to the procedure we mimic here, and the chance that the profiles derived in this way are prone to a $\pm 10-20$ percent uncertainty in the baryon fraction appears high. 

We conclude by noting that the  "clumpy" baryon fraction in all systems is systematically {\it lower} by $\sim 30-40$ percent compared to the baryon fraction
of the whole cluster. This suggest that the clumpiest regions of the ICM are baryon-poor compared to the average ICM. The most likely reason for 
that is that clumps are usually associated with substructures \citep[][]{TO03.1,do09}, that have been subjected to ram-pressure stripping. 
Most of these satellites should indeed lose a sizable part of their gas atmosphere while orbiting within the main halo, ending up with a baryon fraction {\it lower} than the cosmic value. Recent observations of gas-depleted galaxies identified in 300 nearby groups with the Sloan Digital Sky Survey also seem to support this scenario \citep[][]{2012arXiv1207.3924Z}.
This 
mechanism is suggested by numerical simulations \citep[][]{TO03.1,do09}, and 
it is very robust against numerical details \citep[e.g.][]{va11comparison} or the adopted physical implementations \citep[][]{do09}.
However, it seems very hard to probe observationally this
result, because these DM sub-halos need high resolution
data to resolve them both in their gas component
with X-ray and in the total mass distribution through,
e.g., gravitational lensing techniques.

\begin{figure*}
\includegraphics[width=0.32\textwidth]{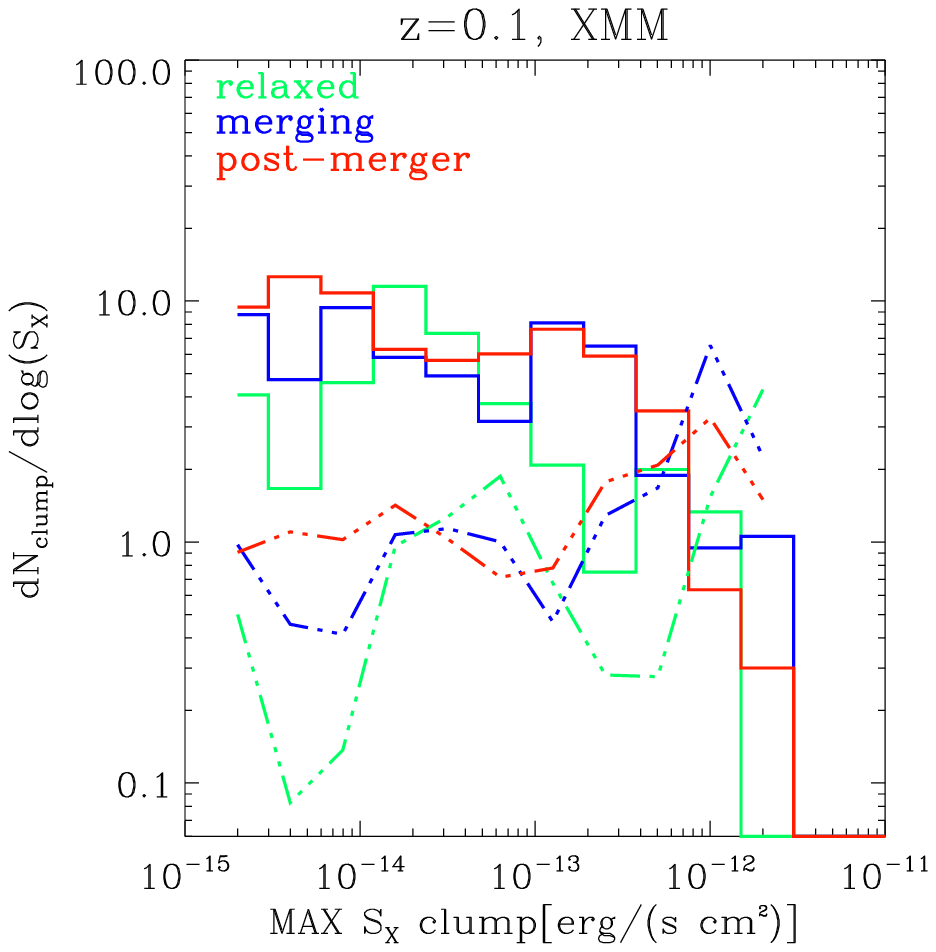}
\includegraphics[width=0.32\textwidth]{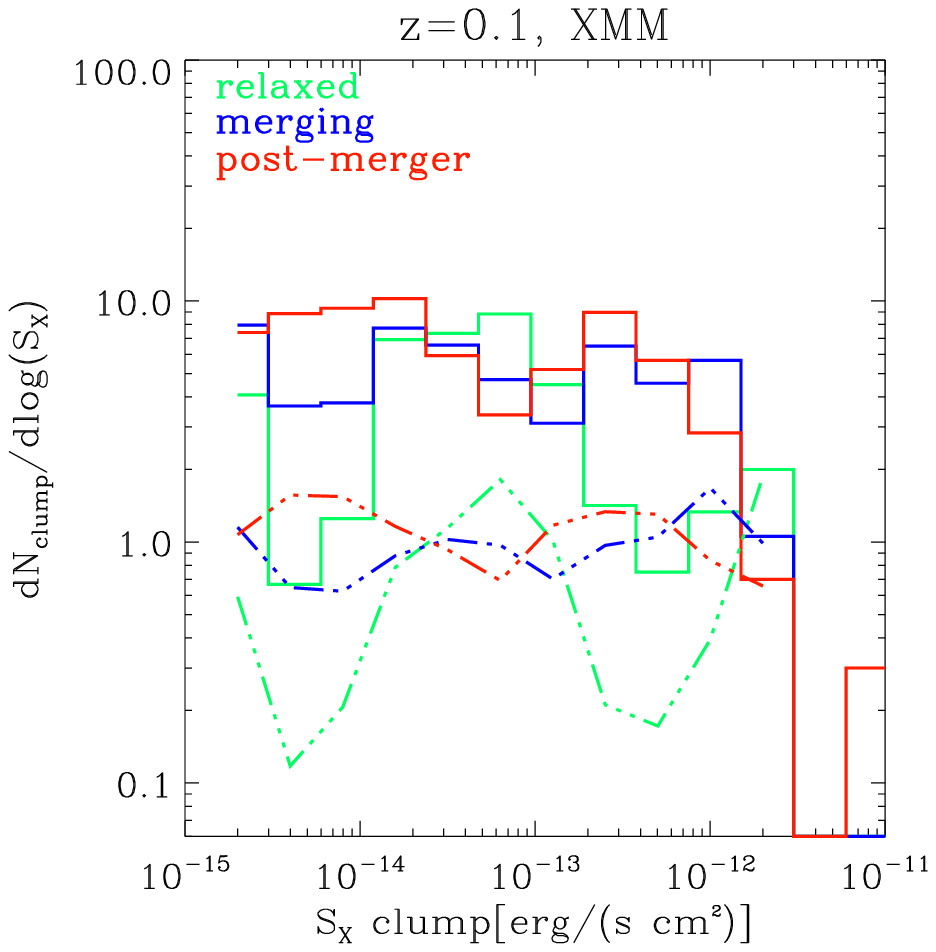}
\includegraphics[width=0.32\textwidth]{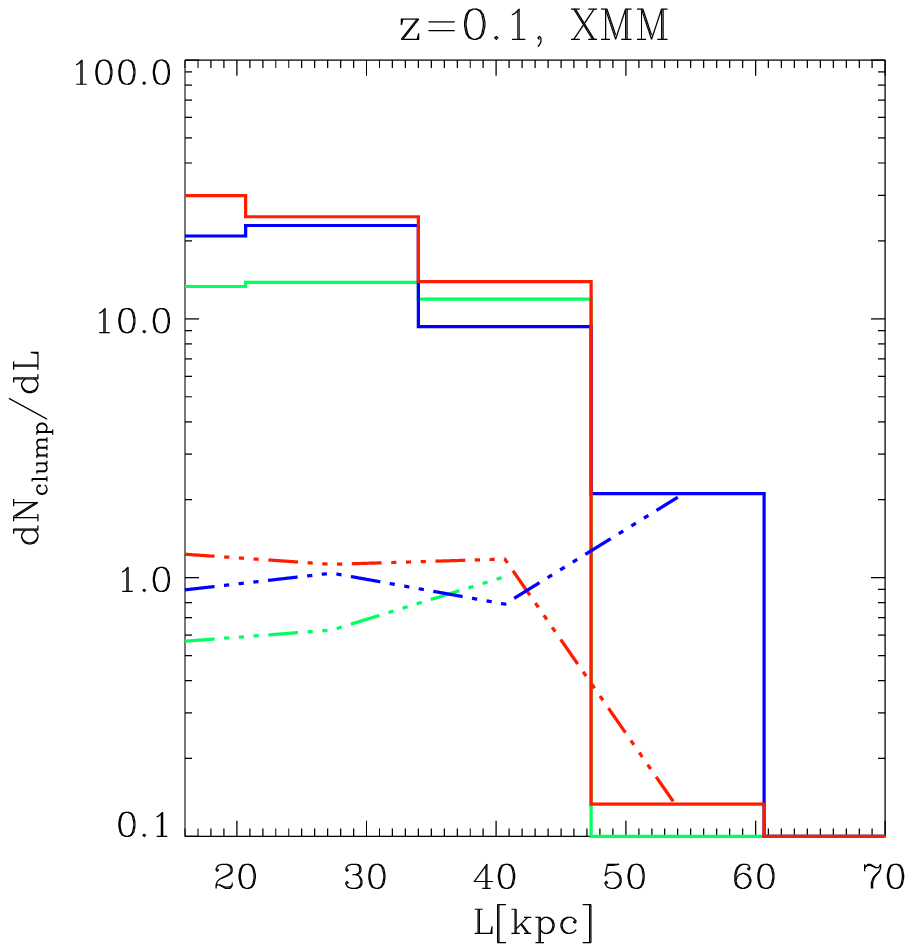}
\includegraphics[width=0.32\textwidth]{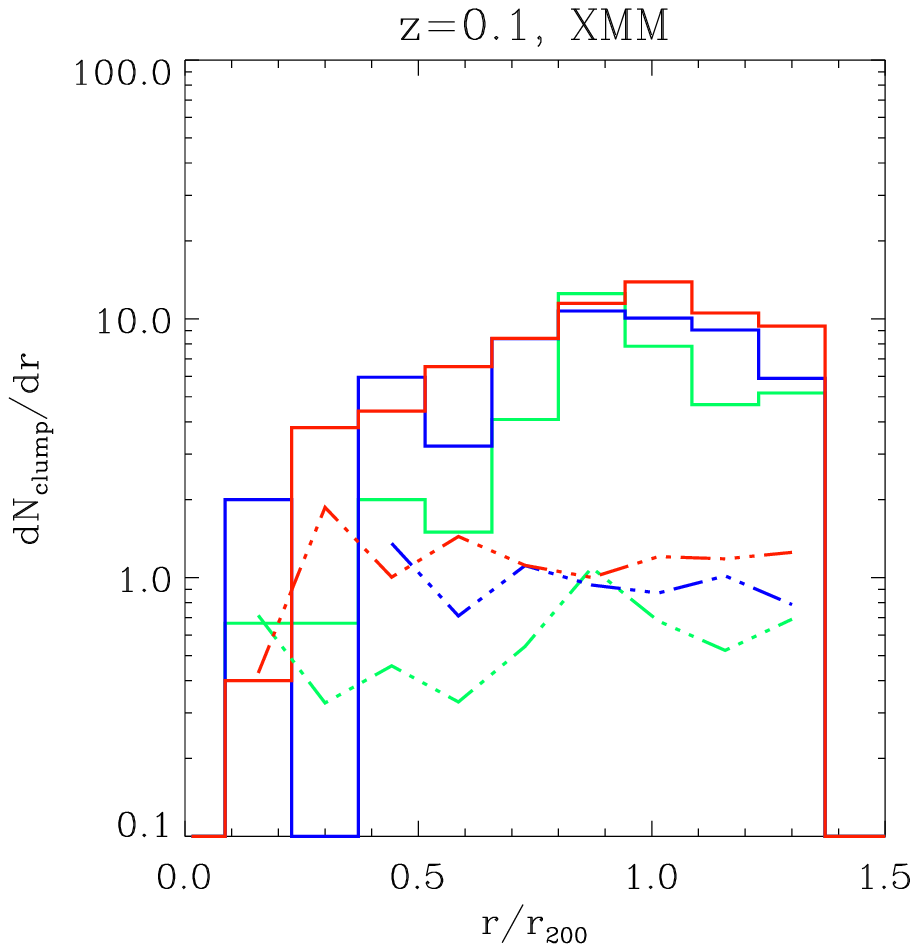}
\includegraphics[width=0.32\textwidth]{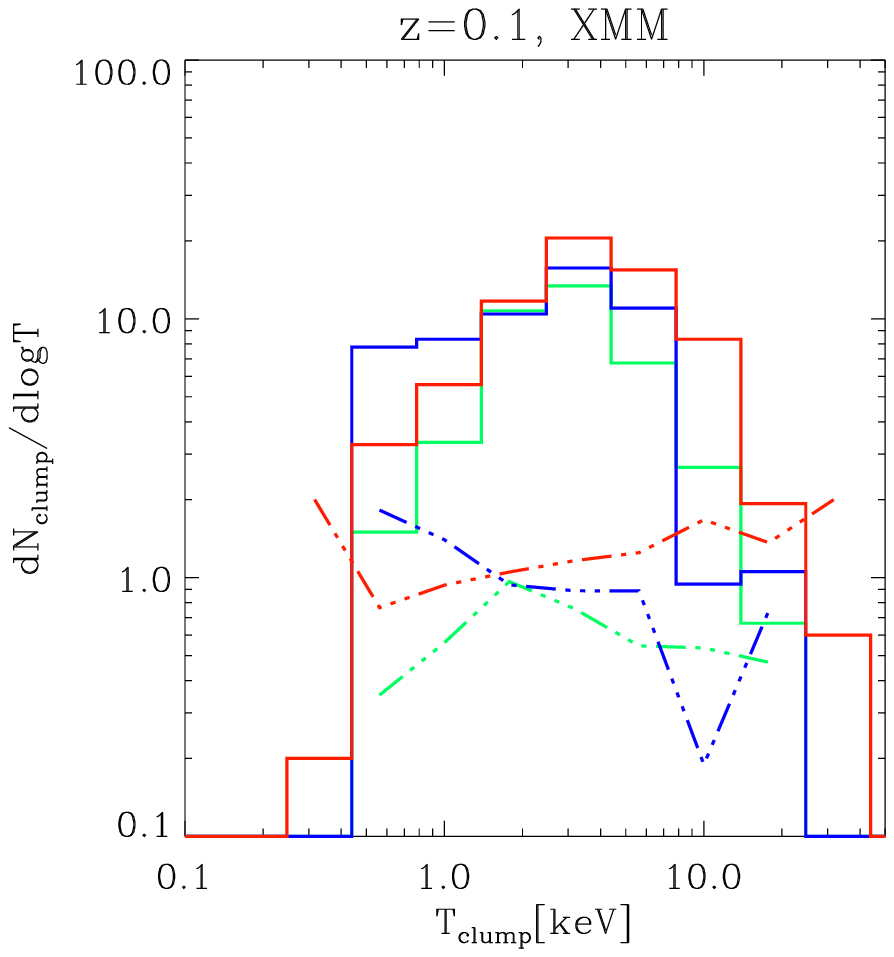}
\includegraphics[width=0.32\textwidth]{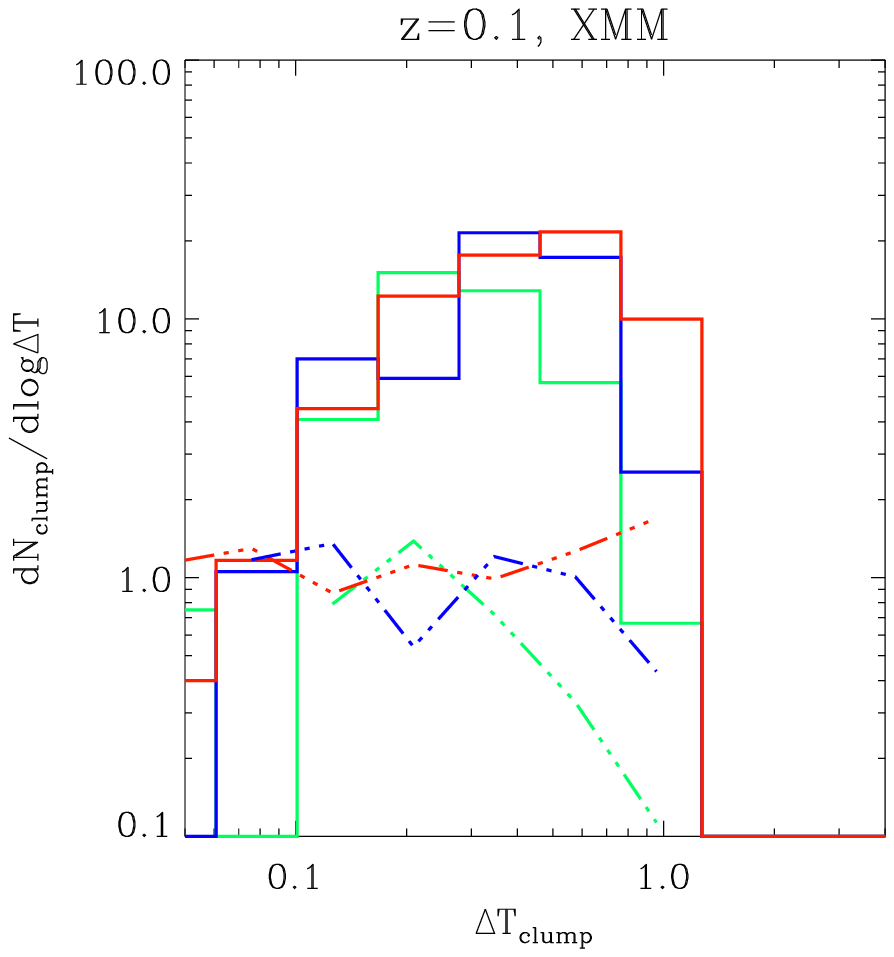}
\caption{Differential distribution functions of clumps detected in our simulated X-ray maps at z=0.1 and assuming $S_{\rm X,low}=2\cdot 10^{-15} \rm erg/(s \cdot cm^2)$ and an effective resolution of $\approx 10''$ (to mimic a deep survey with {\it XMM}). From left to right, we show: the distribution of the maximum and of the total flux of clumps and the distribution of FWHM in clumps in the top panels, the radial distribution, the temperature distribution and the distribution of temperature contrast of clump in the 
bottom panels.
The different colours refer to the dynamical classes in which our sample is divided. 
All distributions have been normalized to the number of objects within each class. The lower dot-dashed lines shows the ratio between clumps in the various dynamical classes, compared to the average population.}
\label{fig:distrib_clumps2}
\end{figure*}

\begin{figure}
\includegraphics[width=0.45\textwidth]{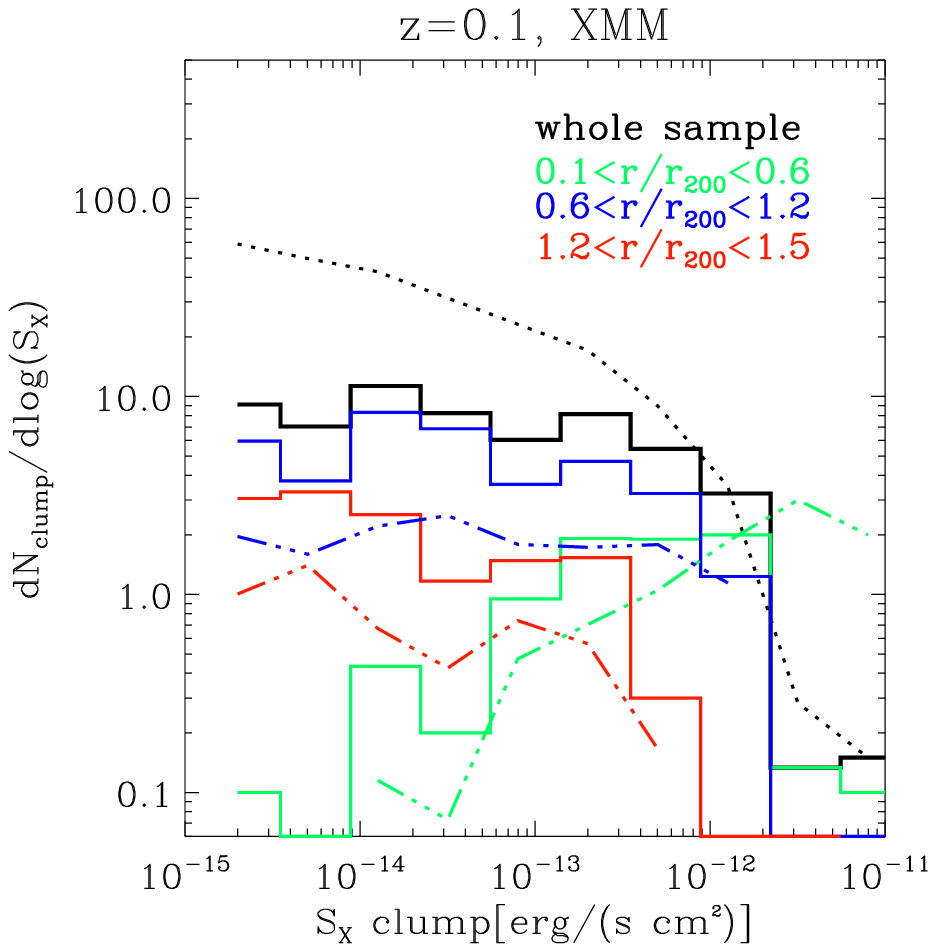}
\includegraphics[width=0.45\textwidth]{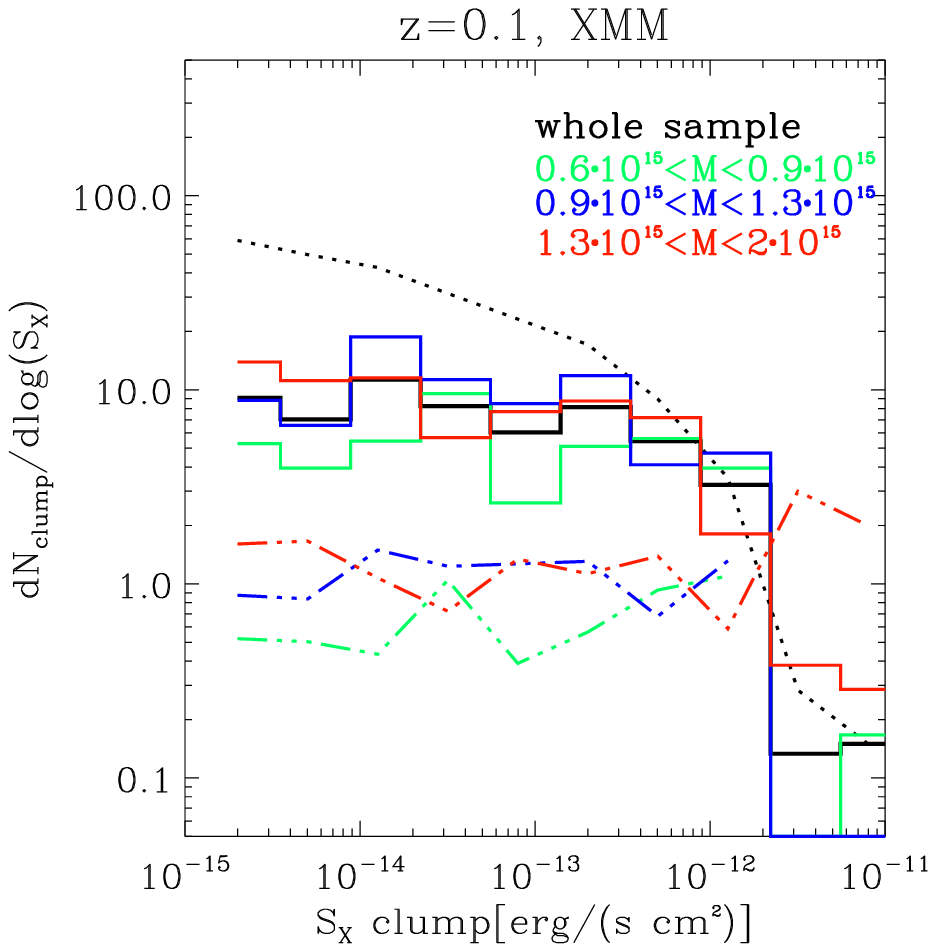}
\caption{Different distribution functions of clumps detected in our simulated X-ray maps at z=0.1 and assuming $S_{\rm X,low}=2 \cdot 10^{-15} \rm erg/(s \cdot cm^2)$ and an effective spatial resolution of $\approx 10''$ (to mimic a deep survey with {\it XMM}). The continuous lines are for the differential distribution, the dashed lines are for the cumulative distributions (only the total is shown). The top panel show the different luminosity functions for three different bin in the projected radii, the bottom panel shows the decompositions of the dataset into three total mass range. The dot-dash lines show the ratio of objects found in each class, compared to the average population.}
\label{fig:distrib_clumps3}
\end{figure}

\subsection{Properties of gas clumps}
\label{subsec:clumps}

From our simulations we can compute the
number of resolved gas clumps in simulated X-ray images.
Based on the clump detection scheme outlined in Sec.\ref{subsubsec:clumps_det}, we produced catalogues of X-ray bright clumps in our images, and we compared the statistics of different 
detection thresholds and spatial resolution, broadly mimicking the realistic performances of
current X-ray satellites. We considered a range of redshifts ($z=0${\footnote {More exactly, we adopted $z=0.023$, corresponding to the luminosity distance of the COMA cluster, $\approx 100 ~\rm Mpc$.}}, $z=0.1$ and $z=0.3$), and
accounted for the effect of cosmological dimming ($\propto (1+z)^4$) and of the reduced linear resolution. We did not generate
photons event files for each specific observational setup, but we
rather assumed a reference sensitivity, $S_{\rm X,low}$, for each
instrument, which was derived from the best cases in the literature. 
The production of more realistic mock X-ray observations from simulated images, calibrated for each instruments, involves a number
of technicalities and problems in modelling \citep[e.g.][]{2005ApJ...618L...1R,heinz10,2012MNRAS.420.3545B}, which
are well beyond the goal of this paper. Our aim, instead, is to assess
the maximum amount of bright clumps that can be imaged
within the most ideal observational condition for each instrument. 
More realistic observational conditions (e.g. a decreasing sensitivity of real instruments moving outwards of the field of view, etc) are
expected to further reduce the real detection ratio of such
clumps (if any), and therefore the results presented in this Section should be considered as upper-limits.

Our set of synthetic observations consists of 60 images (three projections per cluster) for each redshift and instrument. 
In Figure \ref{fig:distrib_clumps1}, we present the luminosity function
of X-ray clumps detected in our maps, assuming three different instrumental setups (we consider here the maximum luminosity per pixel for each clump), representative of realistic X-ray exposures: 10"/pixel and 
$S_{\rm X,low}=2 \cdot 10^{-15} \rm erg/(s \cdot cm^2)$ for {\it XMM}-like observations, 30"/pixel and 
$S_{\rm X,low}=3 \cdot 10^{-14} \rm erg/(s \cdot cm^2)$ for {\it ROSAT}-like observations and 120"/pixel and 
$S_{\rm X,low}=10^{-13} \rm erg/(s \cdot cm^2)$ for {\it Suzaku}-like exposures (for simplicity we considered the [0.5-2] keV energy range in all cases).
In the case of the {\it XMM} "observation" at the two lowest
redshifts, it should be noted that the available spatial resolution
is significantly better than the one probed by our simulated images. In this case, a simple re-binning of our original data to a higher resolution (without the addition of higher resolution modes) has been
adopted.

Based on our results, a significant number
of clumps can be detected in deep exposures with the three instruments:
$\sim 20-30$ per cluster with {\it ROSAT}/{\it Suzaku}, and $\sim 10^2$ per cluster with {\it XMM}. These estimates are based on observations that sample {\it the whole} cluster atmosphere ($1.2 \times 1.2 ~ R_{\rm 200}$).
The evolution with redshift reduces the number of detectable clumps by  $\sim 2-3$, mainly due to the 
effect of the degrading spatial resolution with increasing distance (and not to significant cosmological evolution). At $z=0.3$, only {\it XMM} observations would still detect some clumps ($\sim 30$ per cluster). 
The largest observable flux from clumps, based on these results, is $\sim 5 \cdot 10^{-12} \rm erg/(s \cdot cm^2)$ for {\it Suzaku} and {\it ROSAT} and $\sim 10^{-12} \rm erg/(s \cdot cm^2)$ for {\it XMM} {\footnote{We note that the maximum observable flux here is not completely an intrinsic property of the clumps, but it also slightly depends on the post-processing extraction of clumps in the  projected maps at different resolution, as in Sec.\ref{subsubsec:clumps_det}. For large PSF, indeed, the blending of close clumps can produce a boost in the detectable X-ray flux within the PSF. }}. 

In the following we restrict ourselves to the fiducial case of $z=0.1$ observations with {\it XMM}, and we study the statistical distributions of clumps in more detail. 

\bigskip

In Figure \ref{fig:distrib_clumps2} we show differential distribution functions for the various parameters available for each detected clump:
a) maximum X-ray flux of each identified clump; b) total X-ray flux on the pixel scale; c) projected clump radius (assuming a full-width-half-maximum of the distribution of $S_X${\footnote{The projected clump radius is measured here with an extrapolation, by assuming that
the observed flux from clumps originates from a Gaussian distribution, for which we compute the full-width-half-maximum. This choice is motivated by the fact that in most cases our detected clumps are very close to the resolution limit of our mock observations, and only with 
some extrapolation we can avoid a too coarse binning of this data. For this reason, our measure of the clump radius must be considered only
a first order estimate, for which higher resolution is required in 
order to achieve a better accuracy.}}); d) projected distance from the cluster centre; e) average projected temperature; f) temperature contrast around the clump (estimated by dividing the temperature of the clump by the average surrounding temperature of the ICM, calculated as the average after the exclusion of the clump temperature).

The luminosity distributions of clumps confirm the result of the previous section, and show that post-merger and merging clusters are characterized by a larger number of detectable clumps. They also host on average clumps with a $\sim 2-3$ times larger X-ray flux. Perturbed clusters have a factor $\sim 2-4$ excess of detectable clumps at all radii compared to relaxed clusters, and they host larger clumps. Clumps in post-merger systems are on average by factor $\sim 2$ smaller than in merging systems, likely as an effect of stripping and disruption of clumps during the main merger phase.
The distributions of temperatures and temperature contrasts show that the projected clumps are always colder than the surrounding ICM. This is because the innermost dense region of surviving clumps is shielded from the surrounding ICM, and retains its lower virial
temperature for a few dynamical times \citep[][]{TO03.1}.
For most of our detected clumps, the estimated FWHM is $\leq 50 ~\rm kpc/h$. 
This is close to the cell size in our runs, and raises the problem that the innermost regions of our clumps are probably not yet converged with respect to resolution. We will revisit this issue again in the next section.

In Sec.\ref{fig:distrib_clumps3}, we investigate the dependence of the luminosity function of clumps on the projected radial distance from the centre (top panel) and on the total gas mass.
We found that for all luminosity bins the intermediate annulus ($0.6 \leq R/R_{\rm 200}\leq 1.2$) contains $>70$ percent of detectable clumps, even if its projected volume is about the same as that of the most external annulus. 
On the other hand we detect no significant evolution with the host cluster mass. This is reasonable since we neglect radiative cooling or other processes than can break the self-similarity of our clusters. 

Given the relatively large number of detectable clumps predicted by our simulations, one may wonder whether existing observations with {\it XMM}, {\it Suzaku}, {\it ROSAT} or {\it Chandra} can already be compared with our
results. We will come back to this point in Sec.\ref{sec:conclusions}.

\begin{figure*}
\includegraphics[width=0.95\textwidth]{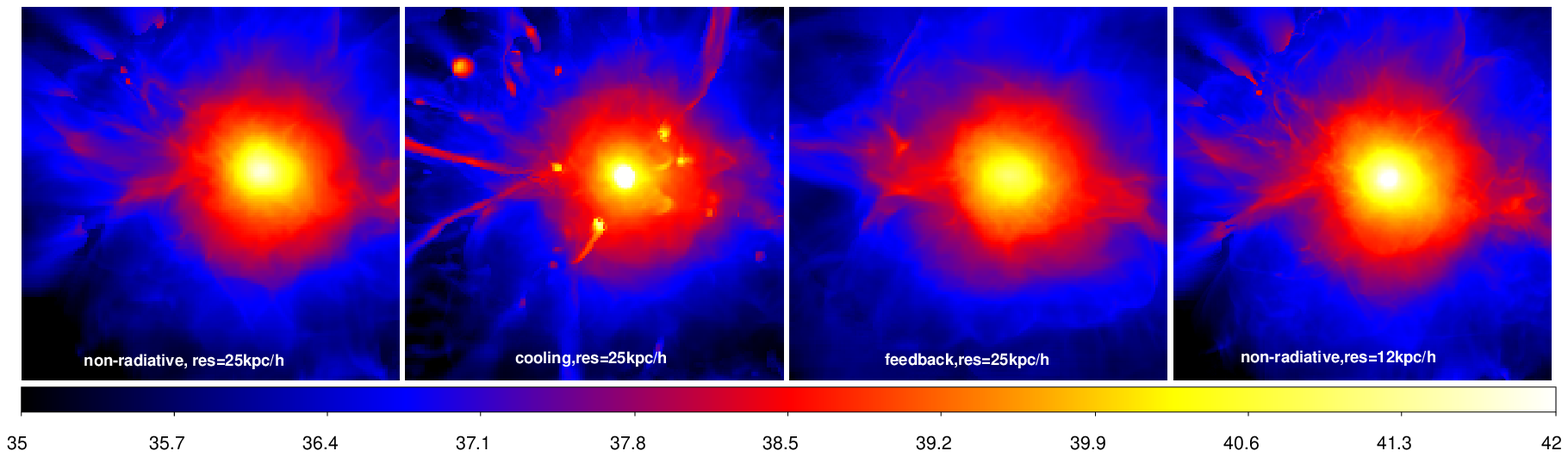}
\includegraphics[width=0.95\textwidth]{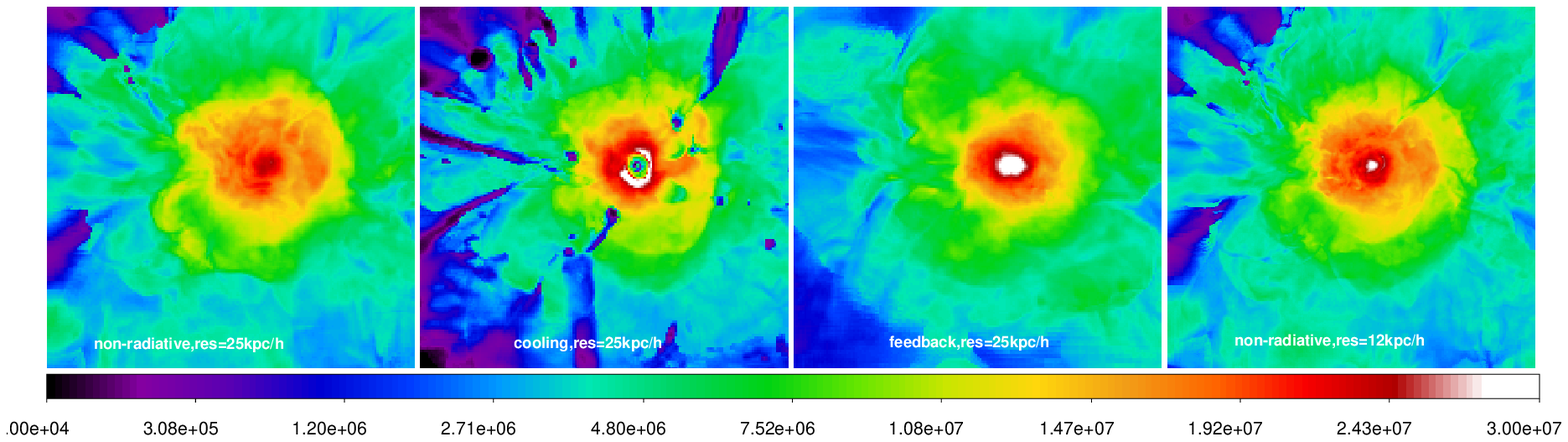}
\caption{Top panels: projected X-ray flux ($log_{\rm 10} \rm L_X$ [erg/s]) of 4 resimulations of a $\sim 3 \cdot 10^{14} M_{\odot}$ employing different physical implementations (see main text for details). Bottom panels: projected spectroscopic-like temperature for the same runs (T in [K]).}
\label{fig:clumping_feedback1}
\end{figure*}

\begin{figure*}
\includegraphics[width=0.9\textwidth]{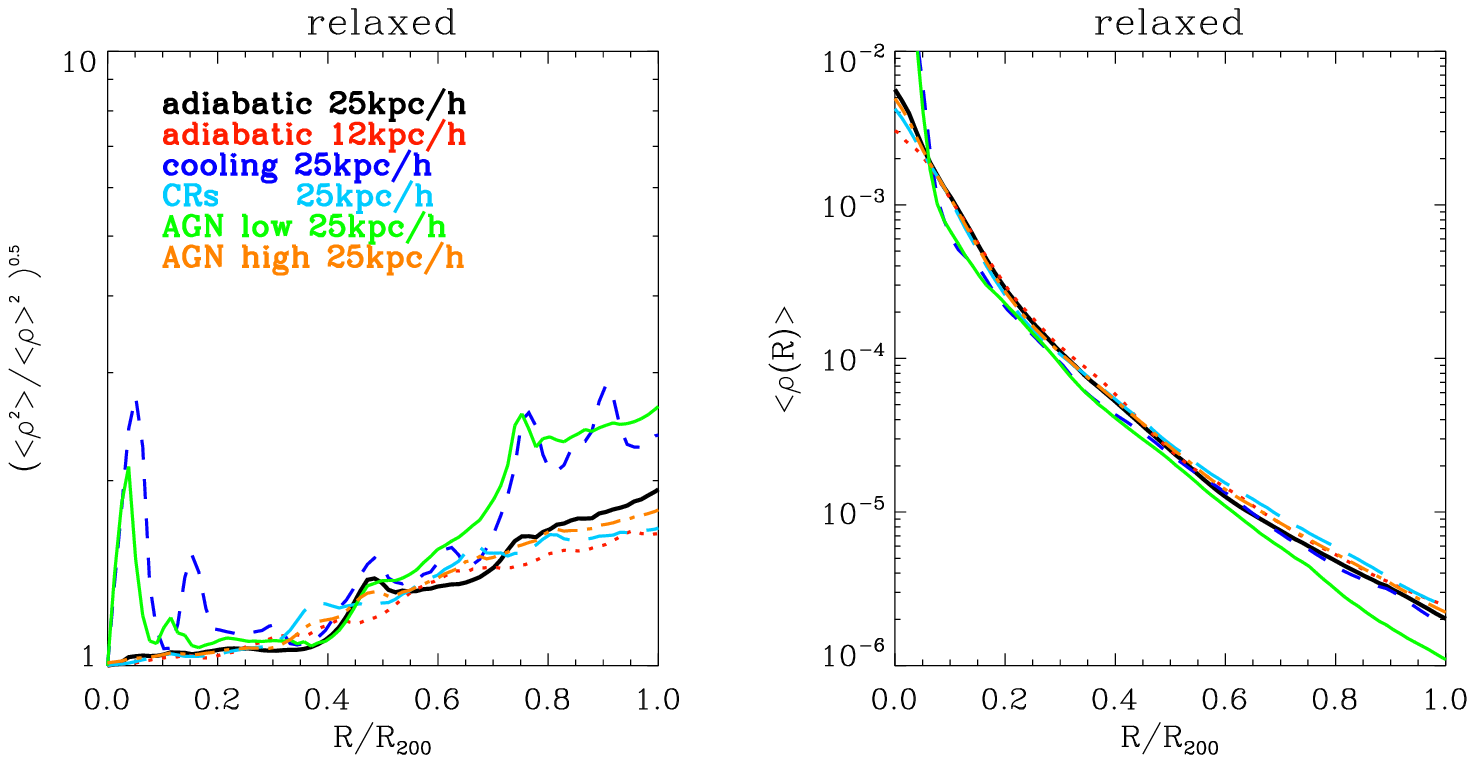}
\includegraphics[width=0.9\textwidth]{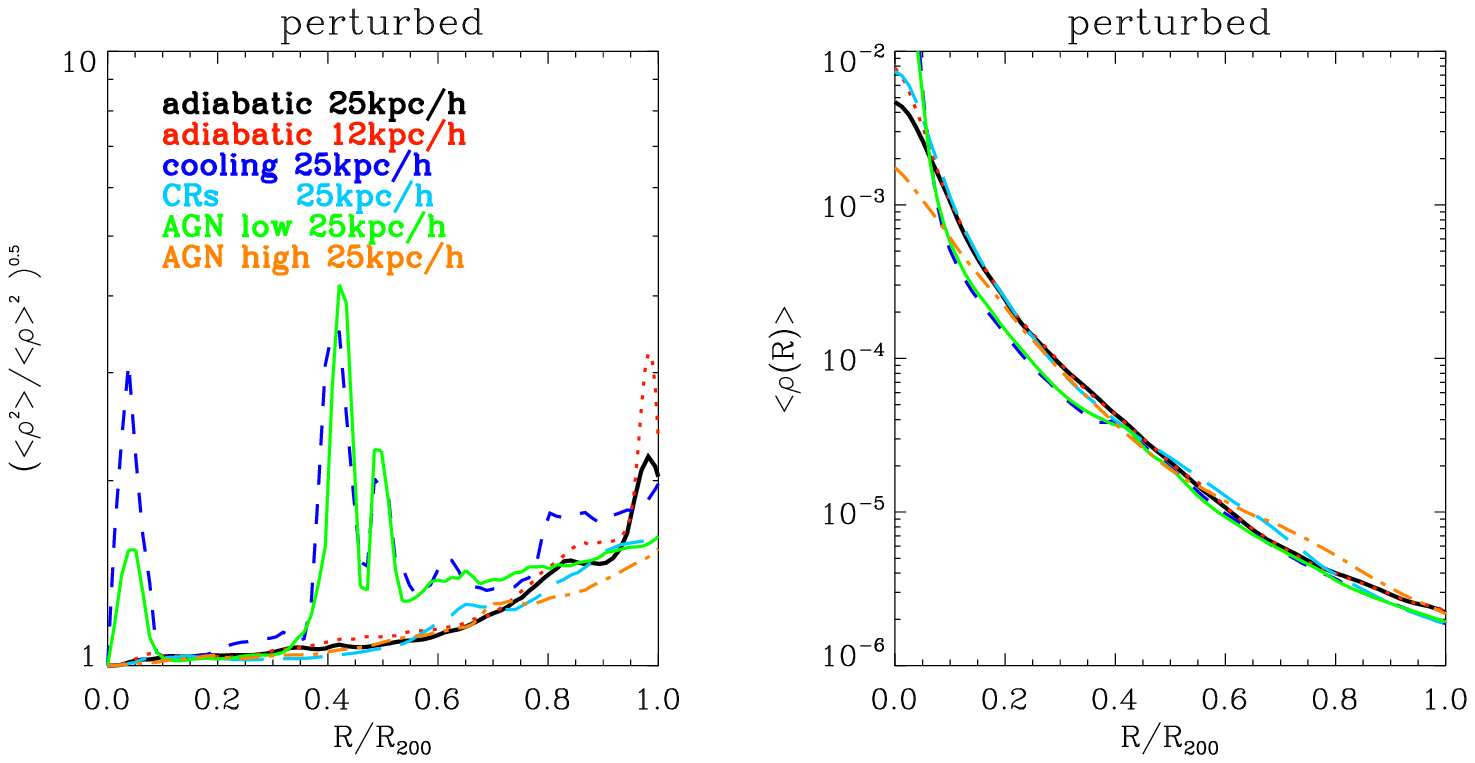}
\caption{Left panels: average profiles of clumping factor for a relaxed and a perturbed (i.e. with an ongoing merger) cluster in the mass range $\sim 2-3 \cdot 10^{14} M_{\odot}$ resimulated with various physical recipes (plus one resimulation at higher resolution).
Right panels: gas density profiles for the corresponding runs on the left panels.}
\label{fig:clumping_feedback2}
\end{figure*}

\begin{figure}
\includegraphics[width=0.45\textwidth]{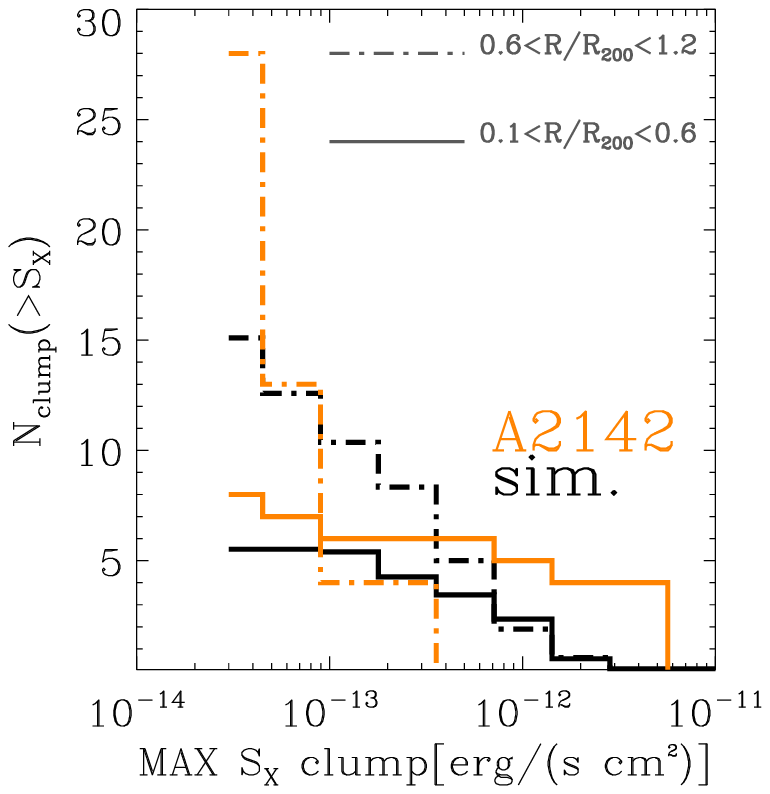}
\includegraphics[width=0.45\textwidth]{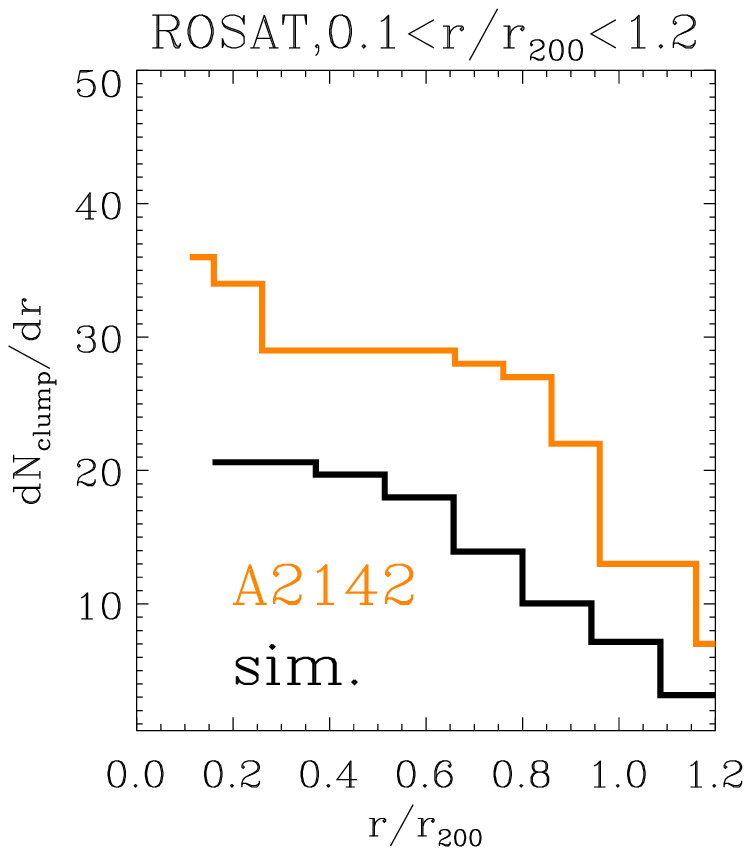}
\caption{Top panel: cumulative distribution of luminosity of simulated clumps 
(in black, the average over the whole dataset is considered) and of point-like
sources in the {\small ROSAT} observation of A2142 \citep[in orange, taken from][]{eckert12}. The solid lines show the distributions inside $0.1 \leq R/R_{200}<0.6$, the dot-dashed lines show the distribution inside $0.6 \leq R/R_{200}<1.2$. Bottom panel: cumulative radial distribution of simulated clumps
and point-like sources for the same datasets.} 
\label{fig:a2142}
\end{figure}

\subsection{Tests with additional physics and at higher resolution}
\label{subsec:test}

Several physical and numerical effects may affect
the amount of gas clumping factor measured in simulations.
For instance, \citet{nala11} recently showed that the number
of clumps in simulations with radiative cooling and feedback from supernovae and star formation is significantly larger
than in non-radiative runs. However, clumps can cool so efficiently that they can eventually cool below X-ray emitting
temperatures \citep[][]{nala11}.
Radiative cooling leads to the formation of
more concentrated high-density clumps in the ICM, while feedback from SN and AGN may tend to wash them out by preventing the cooling
catastrophe and providing the gas of clumps more thermal energy.
The result of this competition between radiative losses and energy input via feedback may depend on the details of implementation. 
Also, the presence of cosmic rays accelerated at cosmological shocks  may yield a different compressibility of the ICM in the cluster outskirts, which
can change significantly the amount of clumping there \citep[][]{scienzo}.

In this section, we assess the uncertainty in our results, using a set of cluster re-simulations with different
setups.
Cluster runs such as the one we discussed in the previous sections are fairly expensive in terms of CPU time ($\sim 3-4 \cdot 10^4$ CPU hours for each cluster), given the number of high-resolution cells generated during the simulation. In order to monitor the effects of different setups we thus opted for a set of smaller clusters, already explored in \citet{va11entropy}. 
We chose a pair of galaxy clusters with final mass of $\sim 2-3 \cdot 10^{14} M_{\odot}$ and two very different dynamical states (one post-merger and one relaxed cluster), and studied in detail the radial distribution of gas clumping factor in each of them. These clusters were re-simulated with the following numerical setups: a) a simple non-radiative run, with maximum resolution of $25 ~\rm kpc/h$; b) a run with pure radiative cooling, same peak resolution; c) a run with thermal feedback from a single, central AGN starting at $z=1$, with a power of $W_{\rm jet} \approx 10^{44} \rm erg/s $ for each injection, with same peak resolution; d) a run with a uniform pre-heating of $100 ~\rm keV cm^2$ (released at $z=10$) followed by a phase of thermal feedback from a central AGN starting at $z=1$ and a power of $W_{\rm jet} \approx 10^{43} \rm erg/s$ for each injection ; e) a run with cosmic-ray injection at shocks, reduced thermalization and pressure feedback, same peak resolution; f) a simple non-radiative run at a higher resolution, $12  ~\rm kpc/h$.

The cases c) and d) are taken from \citet{va11entropy}, where we modelled the effects of a uniform pre-heating background of cosmic gas \citep[following][]{2007ApJ...666..647Y} and of intermittent thermal AGN activity, modelled as bipolar outputs of over-pressurized gas around the peak of cluster gas density. These runs were designed to fit the average thermal properties of nearby galaxy clusters \citep[e.g.][]{cav09} and to quench catastrophic cooling in radiative cluster simulations. We note that the amount of pre-heating and thermal energy release from the "AGN region" were imposed by construction (in order to allow a parameter space exploration) and are not self-consistently derived from a run-time model of super-massive black-holes evolution. In this implementation, we neglect star formation and therefore the energy feedback from supernovae and the gas mass drop-out term related to the star-forming phase is missing, leading to a higher baryon fraction of the hot gas phased compared to more self-consistent recipes \citep[e.g.][and references therein]{borgani08}. In our runs, the baryon fraction of the warm-hot medium is expected to be overestimated by a factor $\sim 2$ in the innermost cluster regions. Given that we excised the innermost $0.1 R_{\rm 200}$ from the analysis of clumps of the previous section, this problem is not expected to be a major one. 

Our model that includes cosmic rays, is taken from \citet{scienzo}, and computes the run-time effects of diffusive shock acceleration \citep[e.g.][]{kj07} in cosmological shocks: reduced thermalization efficiency at shock, injection of CRs energy characterized by the adiabatic index $\Gamma=4/3$, pressure feedback on the thermal gas. This model was developed to perform run-time modifications of the dynamical effect of accelerated relativistic hadrons in large-scale structures. Given the acceleration efficiency of the assumed model \citep[][]{kj07} the pressure ratio between CRs and gas is always small, $\sim 0.05-0.1$ inside $R_{\rm 200}$ of our simulated clusters at $z=0$. 


The feedback models discussed above are an
idealized subset of feedback recipes that have been used by other authors \citep[e.g.][]{2007MNRAS.376.1547C,2007MNRAS.380..877S,gaspari11b}. For the goal
of this paper, they just represent extreme cases of dramatic cooling
or of efficient heating in clusters, while the reality lies most likely  in between.

Figure \ref{fig:clumping_feedback1} shows the projected X-ray maps and spectroscopic-like temperature maps for four re-simulations (a-b-d-f) for the merging cluster. 

The corresponding profiles of average clumping factor and gas density for this cluster (bottom panels) and for the relaxed cluster (top panels) are shown in Fig.\ref{fig:clumping_feedback2}. The trend with the physical implementations is quite clear in both cases.
The inclusion of cooling causes a "cooling catastrophe" and the enhancement of the core gas density in both clusters. At the same time it also triggers the formation of denser and brighter clumps at all radii. A weak amount of thermal feedback from a centrally-located AGN 
 is found to reduce the gas clumping factor only by a moderate
amount, and only very close to the cluster centre ($\leq 0.3 R_{\rm 200}$), whereas it has no effect at larger radii. In these two runs the luminosity of the brightest clumps can be $10-10^2$ times larger than in non-radiative runs.
On the other hand, with a more efficient feedback recipe (early diffuse pre-heating an moderate AGN thermal feedback at low redshift) it seems possible to
quench the cooling catastrophe in both clusters centre, and to reduce at the same time  the gas clumping factor at all radii, making the latter only slightly larger than in the simple non-radiative run. 
The inclusion of CR feedback at cosmological shocks is found not to affect our gas clumping statistics compared to non-radiative runs. This can be explained because in our model the strongest dynamical effect of CRs is found around accretion shocks,
where the energy ratio between injected CR-energy and gas energy can be $\sim 10-20$ percent. However, filaments can enter far into
the main cluster atmosphere avoiding strong accretion shocks, and thus keeping a smaller energy ratio of CRs. Given the strong
role of filaments in enriching the ICM of clumpy material (as shown in Sec.\ref{subsec:clumping}), the fact that the overall
clumping statistics within the cluster is not affected by the injection of CRs at shocks is not surprising.

Finally, we tested the effect of increasing spatial 
resolution on the gas clumping factor in our runs.
On average, a larger resolution in the non-radiative runs is found
not to change the overall gas clumping factor within the cluster,
although assessing its effect on the high density peaks of the
clumps distribution is non trivial. Indeed, the increase in 
resolution may change the ram-pressure stripping history 
of accreted clumps, causing a slightly different orbit of substructures towards the end of runs \citep[e.g.][]{2007MNRAS.380.1399R}. This may lead to time-dependent features
in the gas clumping factor profile, i.e. the higher resolution
run of the perturbed cluster shows a "spike" of enhanced clumping factor at $\sim R_{\rm 200}$ at $z=0$, while no such feature
is detected in the resimulation of the relaxed system. The 
large-scale average trend of the gas clumping factor in both
systems, instead, presents no systematic trend with resolution, suggesting that overall the effect of resolution in non-radiative
runs is not significant for the spatial range explored here. 
However, we caution that some of the parameters of bright gas clumps
derived in this work should be subject to some evolution with
spatial resolution. The small number of clumps found in these two cluster runs is so small that we cannot perform meaningful clumps statistics as in the previous Section.
However, we can already notice that the higher resolution 
increases the brightness of the clumps by a factor $\sim 2-3$. 
In addition, also the typical innermost radius of bright clumps
can be only poorly constrained by our runs, and based on 
Sec.\ref{subsec:clumps} our estimate of a typical $\leq 50 ~\rm kpc/h$
must be considered as an upper limits of their real size, because
the FWHM of most of our clumps is close to our maximum resolution,
and this parameter has likely not yet converged with resolution.


In conclusion, the effect of radiative cooling has a dramatic impact on the properties of clumps in our simulations (as suggested by \citealt{nala11}). However, AGN feedback makes the X-ray flux from clumps very similar to the non-radiative case and the most important results of our analysis (Sec.3.1-3.2) should hold. However, the trend with resolution is more difficult to estimate, given the large numerical cost of simulations at a much higher resolution. 
We therefore must defer the study of gas clumping statistics at much higher resolution to the future. This will also help to understand the number of  {\it unresolved} gas clumps that might contribute to the X-ray emission of nearby clusters.

\section{Discussion and conclusions}
\label{sec:conclusions}

In this paper we used a set of high-resolution cosmological simulations of massive galaxy
clusters to explore the statistics and effects of overdense gas substructures in the intra cluster medium.
While the densest part of such gas substructures may be directly detected in X-ray against the smooth emission of the ICM,
the moderate overdensity associated to substructure increases the gas clumping factor of the ICM, and may produce a (not resolved)
contribution to the X-ray emission profiles of galaxy clusters. 

We analysed the outputs of a sample of 20 massive systems ($\sim 10^{15} M_{\odot}$) simulated at high resolution with the {\small ENZO} code \citep[][]{no07,co11}.
For each object we extracted the profile of the gas clumping factor within the full cluster volume, and within smaller volumes defined by projected sectors from the cluster centre.
The gas clumping factor is found to increase with radius in all clusters,
 in agreement with \citet{nala11}: it is consistent with 1 in the innermost
 cluster regions, and increases to $C_(\rho) \sim 3-5$ approaching the virial radius. Strongly perturbed systems (e.g. systems with an ongoing merger or post-merger systems) are on average characterized by a larger amount of gas clumping factor at all radii.  This enhancement is associated with massive accretion of gas/DM along filaments, which also produces large-scale asymmetries in the radial profiles of clusters. Obtaining an accurate estimate of the enclosed baryon fraction for systems with large-scale accretions can be difficult because of the significant bias of gas clumping factor in the derivation of gas mass, and in the departure of
 real profiles from the NFW-profile, which can bias the estimate of the underlying
 DM mass. In a realistic observation of a narrow sector from the centre of a cluster, the estimate of the enclosed baryon fraction 
can be biased by  $\pm 10$ percent in relaxed systems and by $\pm 20$ percent in systems with large-scale asymmetries (not necessarily associated with major mergers).

In order to investigate the detectability of the high density peaks of the distribution of gas clumping factor,
we produced maps of X-ray emission in the [0.5-2] keV energy band for our clusters at different epochs. We extracted the 
gas clumps present in the images with a filtering technique, selecting all pixels brighter than twice the smoothed X-ray emission of the cluster.
By compiling the luminosity functions and the distribution functions of the most important parameters of the clumps (such has their projected size, temperature and temperature contrast with the surrounding ICM) we studied the evolution of these distributions as a function of the mass and of the dynamical state of with host cluster, and of the epoch of observation.
Summarizing our results on clumps statistics, we find that: a) there is a significant dependence on the number of bright clumps and the dynamical state of the host cluster, with the most perturbed systems hosting on average a factor $\sim 2-10$ more clumps at all radii and luminosities, compared to the most relaxed systems; b) the host cluster mass, on the other hand, does not affect the number of bright clumps (although the investigated mass range is not large, $\sim 0.3$ dex); c)  about a half of detectable bright clumps is located in the radial range $0.6 \leq R/R_{\rm 200} \leq 1.2$ from the projected
cluster centre, while a very small amount of clumps per cluster (order 1) are located inside $<0.6 R/R_{\rm 200}$ ; d) the typical size of most of gas clumps (extrapolated by our data) is $\leq 50$ kpc/h. 
In a preliminary analysis based on a few re-simulations of two simulated clusters with different recipes of CR-physics or AGN feedback, we tested the stability of our results with respect to complex physical models. Radiative cooling
dramatically increases the observable amount of gas clumping \citep[][]{nala11}.
However, the required energy release from AGN seems able to limit at the same
time the cooling flow catastrophe, and the observed clumping factor down to the 
statistics of simpler non-radiative estimates. The increase of resolution does not significantly
change the observed number of detectable clumps or the gas clumping factor of the ICM.
However, resimulations at higher resolution produce
a $\sim 2-3$ larger maximum X-ray flux in the clumps, and therefore in
order to assess the numerical convergence on this parameter further
investigations are likely to be necessary. Also the typical size
of bright clumps cannot yet be robustly constrained by our data,
which only allow us to place an upper limit of the order of $\leq 50 ~\rm kpc/h$.

We conclude by noting that, based on our results of the statistics of bright clumps (Sec.\ref{subsec:clumps}) it seems likely that a given number of them could already
be present in existing X-ray observations by {\small XMM} and by {\small ROSAT} (and, to a much lesser extent given the expected lower statistics, by {\small Suzaku}). However, it is very difficult to distinguish bright gas clumps
from more point-like sources (like galaxies or AGN), given that their expected
size is very small ($<50 ~\rm kpc/h$ ), typically close to the effective PSF of instruments. 
In a real observation, a catalogue of point-like sources is generated
during the data-reduction, and excised from the image. 

In order to perform a first check of the our quantitative results, we examined the catalogue of  point-sources generated in a real {\small ROSAT} observations of 
\citet{eckert12}.
In Fig.\ref{fig:a2142} we show the luminosity and the radial distribution of
point-like sources from the {\small ROSAT} observation of A2142, and the
corresponding simulated statistics obtained by analysing the whole 
simulated dataset at the same resolution and flux threshold of the
real observation. The luminosity distributions are taken for two radial
bins: $0.1 \leq R/R_{200} < 0.6$ and $0.6 \leq R/R_{200} < 1.2$.

The number of point-like sources in the real observation
is $\sim 1.5-2$ times larger than the simulated 
distribution of clumps at all radii. 
The observed luminosity distribution inside the innermost radial bin also shows an excess with respect to the simulated clumps at all luminosities.
In the outer bin, however, the observed luminosity distribution
has a totally different shape with respect to the simulated one. 
It should be noticed than in the outer regions the sensitivity
to point-like sources in the real observation is much degraded.
Therefore, it appears likely that the change in the observed
luminosity distribution function of point-like sources is mainly
driven by instrumental effects. 
We found very similar results in the catalogue of point-sources of the {\small ROSAT} observation of PK0754-191 of \citet{eckert12}.
Based on such small statistics, and given the degrading sensitivity of {\small ROSAT} to point-like sources at large
radii, it is still premature to derive stronger conclusions. The number of expected point-like sources brighter than $\sim 3 \cdot 10^{-14} \rm erg/(s \cdot cm^2)$ in the field is $\sim 20/$deg$^2$, based on the collection of wide field and deep pencil
surveys performed with {\small ROSAT}, {\small Chandra} and {\small XMM} by \citet{2003ApJ...588..696M}. Given the 
projected area of the A2142 observation of \citet{eckert12}, the expected number of point-like field sources is $\sim 6-8$ inside
$R_{\rm 200}$, significantly smaller than both the point-like sources detected in the field of A2142 and the total number
of simulated clumps (that are in the order of $\sim 40$ within $R_{\rm 200}$).
However, one should note that the density of galaxies in a cluster field is much higher than in a typical field, and so is the number of AGNs \citep[see e.g.][]{2007ApJ...664..761M,2012ApJ...754...97H}. Thus, the larger number of detected point sources compared to the expectation from background AGN cannot be directly used as evidence for gas clumping. 

It is possible that a significant fraction of such point-like sources are compact and bright self-gravitating gas clumps in clusters. However, it is
unlikely that they can be identified based on their X-ray morphology, given their small size. 
In the case of high-resolution {\it Chandra} images of nearby clusters, it is possible that some existing data can actually contain
indications of gas clumps. Their detection, however, is made complex
by the strong luminosity contrast required for them to be detected
against the bright innermost cluster atmosphere. Very recently, the
application of sophisticated analysis techniques has
indeed suggested that some gas clumps (with size $\leq 100-200 ~\rm kpc$) may actually be present in nearby {\it Chandra} observations \citep[][]{2012ApJ...746..139A}.

In the next future, cross-correlating with the additional information of temperature in case of available X-ray spectra may unveil the presence of gas clumps in the X-ray images (Fig.7).

\section*{acknowledgements}
We thank our referee for useful comments, that greatly improved the quality of our manuscript.
F.V. and M.B. acknowledge support through grant FOR1254 from the Deutsche Forschungsgemeinschaft (DFG). 
F.V. acknowledges computational resources under the CINECA-INAF 2008-2010 agreement. S.E. and F.V. acknowledge the financial contribution from contracts ASI-INAF I/023/05/0 and I/088/06/0. We acknowledge G. Brunetti and C. Gheller for fruitful collaboration in the production of the simulation runs. Support for this work was provided to A. S.  by NASA through Einstein Postdoctoral Fellowship grant number PF9-00070 awarded by the Chandra X-ray Center, which is operated by the Smithsonian Astrophysical Observatory for NASA under contract NAS8-03060. We thank Marc Audard for kindly providing us his code for the X-ray cooling function.

\bibliographystyle{mnras}
\bibliography{franco}

\begin{thebibliography}{}

\bibitem[\protect\citeauthoryear{{Akamatsu} \& {Kawahara}}{{Akamatsu} \&
  {Kawahara}}{2011}]{2011arXiv1112.3030A}
{Akamatsu} H.,  {Kawahara} H., 2011, ArXiv e-prints

\bibitem[\protect\citeauthoryear{{Allen}, {Schmidt}, \& {Fabian}}{{Allen}
  et~al.}{2002}]{2002MNRAS.334L..11A}
{Allen} S.~W., {Schmidt} R.~W.,  {Fabian} A.~C., 2002, \mnras, 334, L11

\bibitem[\protect\citeauthoryear{{Andrade-Santos}, {Lima Neto}, \&
  {Lagan{\'a}}}{{Andrade-Santos} et~al.}{2012}]{2012ApJ...746..139A}
{Andrade-Santos} F., {Lima Neto} G.~B.,  {Lagan{\'a}} T.~F., 2012, \apj, 746,
  139

\bibitem[\protect\citeauthoryear{{Battaglia} et~al.}{{Battaglia}
  et~al.}{2010}]{2010ApJ...725...91B}
{Battaglia} N., {Bond} J.~R., {Pfrommer} C., {Sievers} J.~L.,  {Sijacki} D.,
  2010, \apj, 725, 91

\bibitem[\protect\citeauthoryear{{Bautz} et~al.}{{Bautz}
  et~al.}{2009}]{2009PASJ...61.1117B}
{Bautz} M.~W. et~al., 2009, \pasj, 61, 1117

\bibitem[\protect\citeauthoryear{{Biffi} et~al.}{{Biffi}
  et~al.}{2012}]{2012MNRAS.420.3545B}
{Biffi} V., {Dolag} K., {B{\"o}hringer} H.,  {Lemson} G., 2012, \mnras, 420,
  3545

\bibitem[\protect\citeauthoryear{{Birnboim} \& {Dekel}}{{Birnboim} \&
  {Dekel}}{2011}]{2011MNRAS.415.2566B}
{Birnboim} Y.,  {Dekel} A., 2011, \mnras, 415, 2566

\bibitem[\protect\citeauthoryear{{Bode} et~al.}{{Bode} et~al.}{2012}]{bode12}
{Bode} P., {Ostriker} J.~P., {Cen} R.,  {Trac} H., 2012, ArXiv e-prints

\bibitem[\protect\citeauthoryear{{B{\"o}hringer} et~al.}{{B{\"o}hringer}
  et~al.}{2010}]{boh10}
{B{\"o}hringer} H. et~al., 2010, \aap, 514, A32

\bibitem[\protect\citeauthoryear{{Borgani} et~al.}{{Borgani}
  et~al.}{2008}]{borgani08}
{Borgani} S., {Diaferio} A., {Dolag} K.,  {Schindler} S., 2008, \ssr, 134, 269

\bibitem[\protect\citeauthoryear{{Burns}, {Skillman}, \& {O'Shea}}{{Burns}
  et~al.}{2010}]{bu10}
{Burns} J.~O., {Skillman} S.~W.,  {O'Shea} B.~W., 2010, \apj, 721, 1105

\bibitem[\protect\citeauthoryear{{Cassano} et~al.}{{Cassano}
  et~al.}{2010}]{cassano10}
{Cassano} R., {Ettori} S., {Giacintucci} S., {Brunetti} G., {Markevitch} M.,
  {Venturi} T.,  {Gitti} M., 2010, \apjl, 721, L82

\bibitem[\protect\citeauthoryear{{Cattaneo} \& {Teyssier}}{{Cattaneo} \&
  {Teyssier}}{2007}]{2007MNRAS.376.1547C}
{Cattaneo} A.,  {Teyssier} R., 2007, \mnras, 376, 1547

\bibitem[\protect\citeauthoryear{{Cavagnolo} et~al.}{{Cavagnolo}
  et~al.}{2009}]{cav09}
{Cavagnolo} K.~W., {Donahue} M., {Voit} G.~M.,  {Sun} M., 2009, \apjs, 182, 12

\bibitem[\protect\citeauthoryear{{Collins} et~al.}{{Collins}
  et~al.}{2010}]{co11}
{Collins} D.~C., {Xu} H., {Norman} M.~L., {Li} H.,  {Li} S., 2010, \apjs, 186,
  308

\bibitem[\protect\citeauthoryear{{Dolag} et~al.}{{Dolag} et~al.}{2009}]{do09}
{Dolag} K., {Borgani} S., {Murante} G.,  {Springel} V., 2009, \mnras, 399, 497

\bibitem[\protect\citeauthoryear{{Dolag}, {Bykov}, \& {Diaferio}}{{Dolag}
  et~al.}{2008}]{do08}
{Dolag} K., {Bykov} A.~M.,  {Diaferio} A., 2008, \ssr, 134, 311

\bibitem[\protect\citeauthoryear{{Dolag} et~al.}{{Dolag} et~al.}{2005}]{do05}
{Dolag} K., {Vazza} F., {Brunetti} G.,  {Tormen} G., 2005, \mnras, 364, 753

\bibitem[\protect\citeauthoryear{{Eckert} et~al.}{{Eckert}
  et~al.}{2011}]{eck11}
{Eckert} D., {Molendi} S., {Gastaldello} F.,  {Rossetti} M., 2011, \aap, 529,
  A133

\bibitem[\protect\citeauthoryear{{Eckert} et~al.}{{Eckert}
  et~al.}{2012}]{eckert12}
{Eckert} D. et~al., 2012, \aap, 541, A57

\bibitem[\protect\citeauthoryear{{Einasto} et~al.}{{Einasto}
  et~al.}{2011}]{2011A&A...531A..75E}
{Einasto} J. et~al., 2011, \aap, 531, A75

\bibitem[\protect\citeauthoryear{{Ettori} \& {Balestra}}{{Ettori} \&
  {Balestra}}{2009}]{2009A&A...496..343E}
{Ettori} S.,  {Balestra} I., 2009, \aap, 496, 343

\bibitem[\protect\citeauthoryear{{Ettori} \& {Molendi}}{{Ettori} \&
  {Molendi}}{2011}]{2011MSAIS..17...47E}
{Ettori} S.,  {Molendi} S., 2011, Memorie della Societa Astronomica Italiana
  Supplementi, 17, 47

\bibitem[\protect\citeauthoryear{{Ettori} et~al.}{{Ettori}
  et~al.}{2009}]{2009A&A...501...61E}
{Ettori} S., {Morandi} A., {Tozzi} P., {Balestra} I., {Borgani} S., {Rosati}
  P., {Lovisari} L.,  {Terenziani} F., 2009, \aap, 501, 61

\bibitem[\protect\citeauthoryear{{Gaspari} et~al.}{{Gaspari}
  et~al.}{2011}]{gaspari11b}
{Gaspari} M., {Brighenti} F., {D'Ercole} A.,  {Melioli} C., 2011, \mnras, 415,
  1549

\bibitem[\protect\citeauthoryear{{George} et~al.}{{George}
  et~al.}{2009}]{geo09}
{George} M.~R., {Fabian} A.~C., {Sanders} J.~S., {Young} A.~J.,  {Russell}
  H.~R., 2009, \mnras, 395, 657

\bibitem[\protect\citeauthoryear{{Haines} et~al.}{{Haines}
  et~al.}{2012}]{2012ApJ...754...97H}
{Haines} C.~P. et~al., 2012, \apj, 754, 97

\bibitem[\protect\citeauthoryear{{Heinz}, {Br{\"u}ggen}, \& {Morsony}}{{Heinz}
  et~al.}{2010}]{heinz10}
{Heinz} S., {Br{\"u}ggen} M.,  {Morsony} B., 2010, \apj, 708, 462

\bibitem[\protect\citeauthoryear{{Hoshino} et~al.}{{Hoshino}
  et~al.}{2010}]{2010PASJ...62..371H}
{Hoshino} A. et~al., 2010, \pasj, 62, 371

\bibitem[\protect\citeauthoryear{{Humphrey} et~al.}{{Humphrey}
  et~al.}{2012}]{2012ApJ...748...11H}
{Humphrey} P.~J., {Buote} D.~A., {Brighenti} F., {Flohic} H.~M.~L.~G.,
  {Gastaldello} F.,  {Mathews} W.~G., 2012, \apj, 748, 11

\bibitem[\protect\citeauthoryear{{Iapichino} et~al.}{{Iapichino}
  et~al.}{2011}]{iapichino11}
{Iapichino} L., {Schmidt} W., {Niemeyer} J.~C.,  {Merklein} J., 2011, \mnras,
  414, 2297

\bibitem[\protect\citeauthoryear{{Kang} \& {Jones}}{{Kang} \&
  {Jones}}{2007}]{kj07}
{Kang} H.,  {Jones} T.~W., 2007, Astroparticle Physics, 28, 232

\bibitem[\protect\citeauthoryear{{Kawaharada} et~al.}{{Kawaharada}
  et~al.}{2010}]{2010ApJ...714..423K}
{Kawaharada} M. et~al., 2010, \apj, 714, 423

\bibitem[\protect\citeauthoryear{{Lapi}, {Fusco-Femiano}, \&
  {Cavaliere}}{{Lapi} et~al.}{2010}]{lapi10}
{Lapi} A., {Fusco-Femiano} R.,  {Cavaliere} A., 2010, \aap, 516, A34

\bibitem[\protect\citeauthoryear{{Lau}, {Kravtsov}, \& {Nagai}}{{Lau}
  et~al.}{2009}]{lau09}
{Lau} E.~T., {Kravtsov} A.~V.,  {Nagai} D., 2009, \apj, 705, 1129

\bibitem[\protect\citeauthoryear{{Leccardi} \& {Molendi}}{{Leccardi} \&
  {Molendi}}{2008}]{2008A&A...487..461L}
{Leccardi} A.,  {Molendi} S., 2008, \aap, 487, 461

\bibitem[\protect\citeauthoryear{{Martini}, {Mulchaey}, \& {Kelson}}{{Martini}
  et~al.}{2007}]{2007ApJ...664..761M}
{Martini} P., {Mulchaey} J.~S.,  {Kelson} D.~D., 2007, \apj, 664, 761

\bibitem[\protect\citeauthoryear{{Mathews} \& {Guo}}{{Mathews} \&
  {Guo}}{2011}]{ma11}
{Mathews} W.~G.,  {Guo} F., 2011, \apj, 736, 6

\bibitem[\protect\citeauthoryear{{Mathiesen}, {Evrard}, \& {Mohr}}{{Mathiesen}
  et~al.}{1999}]{1999ApJ...520L..21M}
{Mathiesen} B., {Evrard} A.~E.,  {Mohr} J.~J., 1999, \apjl, 520, L21

\bibitem[\protect\citeauthoryear{{Moretti} et~al.}{{Moretti}
  et~al.}{2003}]{2003ApJ...588..696M}
{Moretti} A., {Campana} S., {Lazzati} D.,  {Tagliaferri} G., 2003, \apj, 588,
  696

\bibitem[\protect\citeauthoryear{{Nagai} \& {Lau}}{{Nagai} \&
  {Lau}}{2011}]{nala11}
{Nagai} D.,  {Lau} E.~T., 2011, \apjl, 731, L10

\bibitem[\protect\citeauthoryear{Navarro, Frenk, \& White}{Navarro
  et~al.}{1996}]{NA96.1}
Navarro J., Frenk C.,  White S., 1996, ApJ, 462, 563

\bibitem[\protect\citeauthoryear{{Norman} et~al.}{{Norman} et~al.}{2007}]{no07}
{Norman} M.~L., {Bryan} G.~L., {Harkness} R.,  {Bordner} J.~a., 2007, ArXiv
  e-prints, 705

\bibitem[\protect\citeauthoryear{{Rasia} et~al.}{{Rasia}
  et~al.}{2005}]{2005ApJ...618L...1R}
{Rasia} E., {Mazzotta} P., {Borgani} S., {Moscardini} L., {Dolag} K., {Tormen}
  G., {Diaferio} A.,  {Murante} G., 2005, \apjl, 618, L1

\bibitem[\protect\citeauthoryear{{Rasia}, {Tormen}, \& {Moscardini}}{{Rasia}
  et~al.}{2004}]{rasia04}
{Rasia} E., {Tormen} G.,  {Moscardini} L., 2004, \mnras, 351, 237

\bibitem[\protect\citeauthoryear{{Reiprich} et~al.}{{Reiprich}
  et~al.}{2009}]{2009A&A...501..899R}
{Reiprich} T.~H. et~al., 2009, \aap, 501, 899

\bibitem[\protect\citeauthoryear{{Roediger} \& {Br{\"u}ggen}}{{Roediger} \&
  {Br{\"u}ggen}}{2007}]{2007MNRAS.380.1399R}
{Roediger} E.,  {Br{\"u}ggen} M., 2007, \mnras, 380, 1399

\bibitem[\protect\citeauthoryear{{Roncarelli} et~al.}{{Roncarelli}
  et~al.}{2006}]{ro06}
{Roncarelli} M., {Ettori} S., {Dolag} K., {Moscardini} L., {Borgani} S.,
  {Murante} G., 2006, \mnras, 373, 1339

\bibitem[\protect\citeauthoryear{{Sato} et~al.}{{Sato}
  et~al.}{2012}]{2012arXiv1203.1700S}
{Sato} T. et~al., 2012, ArXiv e-prints

\bibitem[\protect\citeauthoryear{{Sijacki} et~al.}{{Sijacki}
  et~al.}{2007}]{2007MNRAS.380..877S}
{Sijacki} D., {Springel} V., {Di Matteo} T.,  {Hernquist} L., 2007, \mnras,
  380, 877

\bibitem[\protect\citeauthoryear{{Simionescu} et~al.}{{Simionescu}
  et~al.}{2011}]{si11}
{Simionescu} A. et~al., 2011, Science, 331, 1576

\bibitem[\protect\citeauthoryear{{Smith} et~al.}{{Smith}
  et~al.}{2001}]{2001ApJ...556L..91S}
{Smith} R.~K., {Brickhouse} N.~S., {Liedahl} D.~A.,  {Raymond} J.~C., 2001,
  \apjl, 556, L91

\bibitem[\protect\citeauthoryear{Tormen, Bouchet, \& White}{Tormen
  et~al.}{1997}]{TO97.2}
Tormen G., Bouchet F.,  White S., 1997, MNRAS, 286, 865

\bibitem[\protect\citeauthoryear{Tormen, Moscardini, \& Yoshida}{Tormen
  et~al.}{2003}]{TO03.1}
Tormen G., Moscardini L.,  Yoshida N., 2003, MNRAS, submitted; preprint
  astro-ph/0304375

\bibitem[\protect\citeauthoryear{{Urban} et~al.}{{Urban} et~al.}{2011}]{ur11}
{Urban} O., {Werner} N., {Simionescu} A., {Allen} S.~W.,  {B{\"o}hringer} H.,
  2011, \mnras, 414, 2101

\bibitem[\protect\citeauthoryear{{Vazza}}{{Vazza}}{2011a}]{va11nice}
{Vazza} F., 2011a, ArXiv e-prints

\bibitem[\protect\citeauthoryear{{Vazza}}{{Vazza}}{2011b}]{va11entropy}
{Vazza} F., 2011b, \mnras, 410, 461

\bibitem[\protect\citeauthoryear{{Vazza} et~al.}{{Vazza}
  et~al.}{2012}]{scienzo}
{Vazza} F., {Br{\"u}ggen} M., {Gheller} C.,  {Brunetti} G., 2012, \mnras, 2518

\bibitem[\protect\citeauthoryear{{Vazza}, {Brunetti}, \& {Gheller}}{{Vazza}
  et~al.}{2009}]{va09shocks}
{Vazza} F., {Brunetti} G.,  {Gheller} C., 2009, \mnras, 395, 1333

\bibitem[\protect\citeauthoryear{{Vazza} et~al.}{{Vazza} et~al.}{2010}]{va10kp}
{Vazza} F., {Brunetti} G., {Gheller} C.,  {Brunino} R., 2010, \na, 15, 695

\bibitem[\protect\citeauthoryear{{Vazza} et~al.}{{Vazza}
  et~al.}{2011a}]{va11turbo}
{Vazza} F., {Brunetti} G., {Gheller} C., {Brunino} R.,  {Br{\"u}ggen} M.,
  2011a, \aap, 529, A17

\bibitem[\protect\citeauthoryear{{Vazza} et~al.}{{Vazza}
  et~al.}{2011b}]{va11comparison}
{Vazza} F., {Dolag} K., {Ryu} D., {Brunetti} G., {Gheller} C., {Kang} H.,
  {Pfrommer} C., 2011b, \mnras, 418, 960

\bibitem[\protect\citeauthoryear{{Vazza} et~al.}{{Vazza}
  et~al.}{2011c}]{va11scatter}
{Vazza} F., {Roncarelli} M., {Ettori} S.,  {Dolag} K., 2011c, \mnras, 413, 2305

\bibitem[\protect\citeauthoryear{{Vazza} et~al.}{{Vazza}
  et~al.}{2006}]{2006MNRAS.369L..14V}
{Vazza} F., {Tormen} G., {Cassano} R., {Brunetti} G.,  {Dolag} K., 2006,
  \mnras, 369, L14

\bibitem[\protect\citeauthoryear{{Vikhlinin} et~al.}{{Vikhlinin}
  et~al.}{2006}]{2006ApJ...640..691V}
{Vikhlinin} A., {Kravtsov} A., {Forman} W., {Jones} C., {Markevitch} M.,
  {Murray} S.~S.,  {Van Speybroeck} L., 2006, \apj, 640, 691

\bibitem[\protect\citeauthoryear{{Walker} et~al.}{{Walker}
  et~al.}{2012}]{2012MNRAS.tmp.2785W}
{Walker} S.~A., {Fabian} A.~C., {Sanders} J.~S., {George} M.~R.,  {Tawara} Y.,
  2012, \mnras, 2785

\bibitem[\protect\citeauthoryear{{Younger} \& {Bryan}}{{Younger} \&
  {Bryan}}{2007}]{2007ApJ...666..647Y}
{Younger} J.~D.,  {Bryan} G.~L., 2007, \apj, 666, 647

\bibitem[\protect\citeauthoryear{{Zhang} et~al.}{{Zhang}
  et~al.}{2012}]{2012arXiv1207.3924Z}
{Zhang} W., {Li} C., {Kauffmann} G.,  {Xiao} T., 2012, ArXiv e-prints

\bibitem[\protect\citeauthoryear{{Zhuravleva} et~al.}{{Zhuravleva}
  et~al.}{2012}]{2012arXiv1210.6706Z}
{Zhuravleva} I., {Churazov} E., {Kravtsov} A., {Lau} E.~T., {Nagai} D.,
  {Sunyaev} R., 2012, ArXiv e-prints

\end{thebibliography}

\end{document}